\documentclass[a4paper, 11pt]{article}

\usepackage[latin1]{inputenc}
\usepackage[T1]{fontenc}
\usepackage[sc]{mathpazo} 
\usepackage{courier}
\usepackage{float}
\usepackage[intlimits,fleqn]{amsmath}
\usepackage{amsfonts}
\usepackage{amsbsy}
\usepackage{amssymb,dcolumn,exscale}
\usepackage[T1]{fontenc}
\usepackage{booktabs}
\usepackage{varioref}
\usepackage[small]{caption}
\usepackage[width=16.6cm, left=2.2cm,top=2.5cm,height=23.7cm]{geometry}
\usepackage{cite}
\usepackage{graphicx,color}
\usepackage{epsf}
\usepackage{pstricks,booktabs,multirow}

\graphicspath{{./pics/}}
\DeclareGraphicsExtensions{.eps,.ps}

\newcommand{\Cf}{C_F}
\newcommand{\Ca}{C_A}
\newcommand{\nf}{n_f}
\newcommand{\as}{\alpha_s}

\newcommand{\half}{\ensuremath{\tfrac{1}{2}}}
\newcommand{\mq}{m_{\tilde{q}}}
\newcommand{\mg}{m_{\tilde{g}}}
\newcommand{\mstop}[1][{}]{m_{\tilde{t}_{#1}}}
\newcommand{\shat}{\hat{s}}
\newcommand{\qqbar}{\ensuremath{q \bar{q}}}
\newcommand{\stoppair}{\ensuremath{\tilde{t}_i\,\tilde{t}_i^{\ast}}}
\newcommand{\dif}{\ensuremath{\mathrm{d}}}

\newcommand{\GeV}{\ensuremath{\,\mathrm{GeV}}}
\newcommand{\TeV}{\ensuremath{\,\mathrm{TeV}}}
\newcommand{\fb}{\ensuremath{\,\mathrm{fb}}}
\newcommand{\pb}{\ensuremath{\,\mathrm{pb}}}
\newcommand{\lnbeta}[1][{}]{\ensuremath{\ln^{#1}\beta}}
\newcommand{\lnzwei}[1][{}]{\ln^{#1} 2}
\newcommand{\lnNt}[1][{}]{\ln^{#1}\widetilde{N}}
\newcommand{\Ln}[1][{}]{\ensuremath{\ln^{#1}\Big(\frac{\mu^2}{\mstop^2}\Big)}}

\def\shat{{\hat s}}
\def\muf{{\mu^{}_f}}
\def\mufs{{\mu^{\,2}_f}}
\def\mur{{\mu^{}_r}}
\def\murs{{\mu^{\,2}_r}}

\DeclareMathOperator*{\Li}{Li_2}

\allowdisplaybreaks[3]

\begin{document}
\begin{titlepage}
\noindent
\begin{center}
\Large{\bf
Threshold Improved QCD Corrections for Stop-Antistop production at Hadron colliders
}\\
\vspace{1.5cm}
\large
U.~Langenfeld\\[10mm]
\normalsize
{\it 
Humboldt-Universit\"at zu Berlin\\
\vspace{0.1cm}
Newtonstra\ss e 15, D--12489 Berlin, Germany\\[2mm]
and\\[2mm]
Julius-Maximilians-Universit\"at W\"urzburg\\[1mm]
Am Hubland, D--97074 W\"urzburg, Germany \\[2mm]
\texttt{\footnotesize{ulangenfeld@physik.uni-wuerzburg.de}}
}\\
\vspace{2.3cm}

\large
{\bf Abstract}
\vspace{-0.2cm}
\end{center}
I present improved predictions for the total hadronic cross section
of stop-antistop production at hadron colliders including 
next-to-next-to-leading-order threshold corrections
and approximated Coulomb corrections. 
The results are based on soft corrections, which are 
logarithmically enhanced near threshold.
I present analytic formulas for the NNLO scaling functions
at threshold and explicit numbers for the total hadronic cross sections
for the Tevatron and the LHC. 
Finally I discuss the systematic error, the scale uncertainty and the
PDF error of the hadronic cross section.

\vfill
\end{titlepage}

\newpage
\section{Introduction}
\indent
The Minimal Supersymmetric Standard model (MSSM) is an attractive 
extension \cite{Haber:1984rc,Nilles:1983ge} of the very successful Standard Model.
One property of this theory is its rich spectrum of new heavy particles
which might be discovered at the LHC if they are lighter than $\approx 2\TeV$.
Searches for superymmetric particles are performed at the Tevatron and the LHC.
No superpartners have been discovered so far.
In the mSugra framework of the Minimal Supersymmetric Standard Model (MSSM),
the lightest stop squark, 
one of the scalar supersymmetric partner particle of the top quark, 
is assumed to be the lightest supersymmetric coloured particle, 
lighter than the other scalar quarks.
This is due to large nondiagonal elements in the stop mixing matrix,
see f. e. the review~\cite{Aitchison:2007fn}.
The lightest stop squark might be the first coloured SUSY particle to be discovered. 
The cross section delivers information about the stop mass or, 
if the mass of the stop squark is roughly known from elsewhere, information about 
its spin\cite{Kane:2008kw}.
If these particles cannot be discovered at the Tevatron or LHC, precise cross sections
help to improve mass exclusion limits.

\begin{figure}[t]
\centering
\scalebox{0.65}{\includegraphics{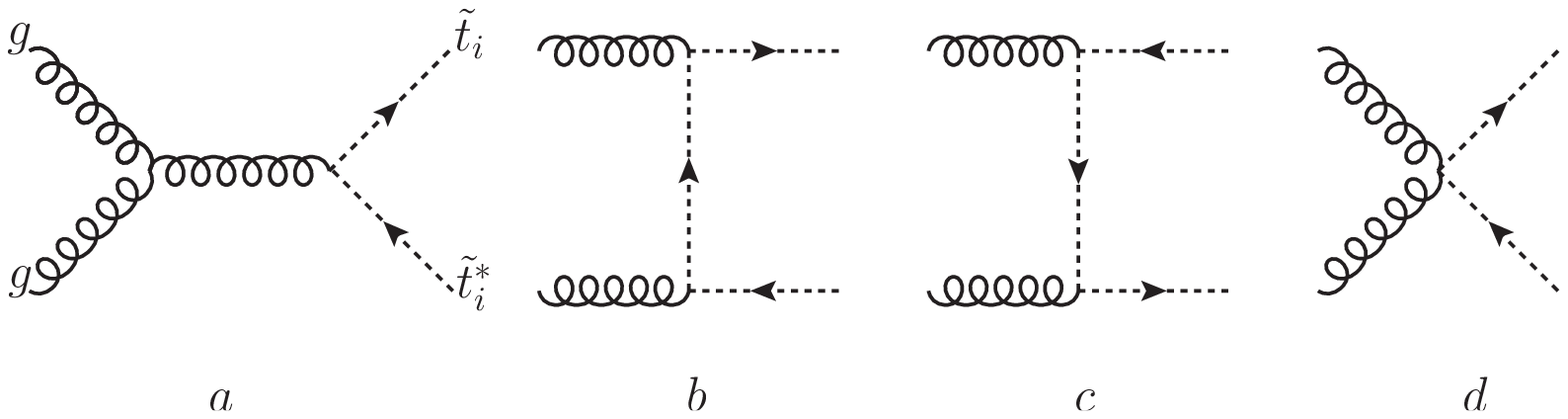}}
%
\scalebox{0.46}{\includegraphics{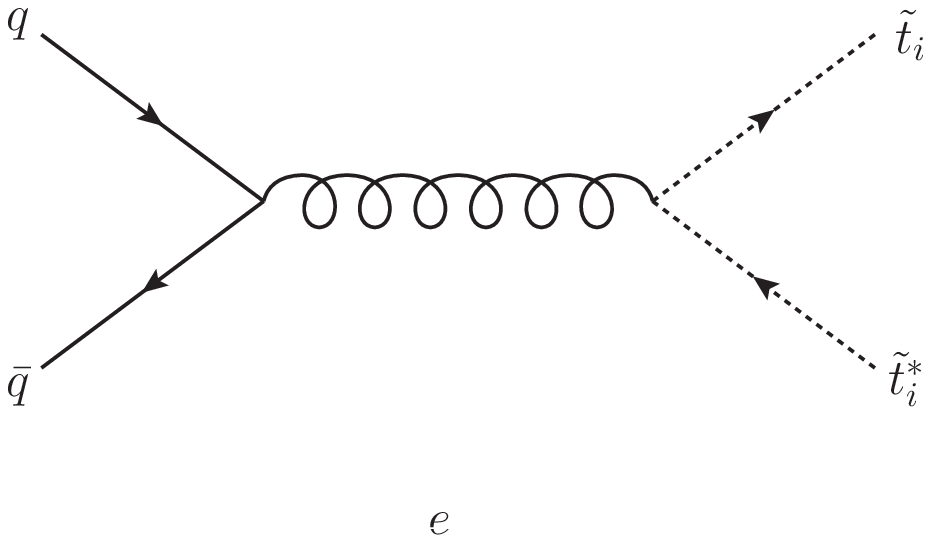}}\vspace*{3mm}
\caption{\small{LO production of a \stoppair~pair via $gg$ annihilation
	(diagrams a-d)
    and \qqbar~annihilation (diagram e).}}
\label{fig:ggbarchannel}
\end{figure}

In this paper, I study the hadroproduction of stop-antistop-pairs
\begin{eqnarray}
 pp/p\bar{p} &\rightarrow& \tilde{t}_i \tilde{t}_i^\ast X,\enspace i=1,\,2,
\end{eqnarray}
with its partonic subprocesses
\begin{eqnarray}
 gg\rightarrow \tilde{t}_i \tilde{t}_i^\ast\enspace \text{and}\enspace
\qqbar \rightarrow\tilde{t}_i\tilde{t}_i^\ast,\enspace q = u,d,c,s,b,
\end{eqnarray}
including NNLO threshold contributions.
The relevant leading order (LO) Feynman diagrams are shown in Fig.~\ref{fig:ggbarchannel}.
The production of mixed stop pairs $\tilde{t}_1\tilde{t}_2^\ast$ or $\tilde{t}_2\tilde{t}_1^\ast$
starts at next to leading order (NLO)~\cite{Beenakker:1997ut} 
and is therefore suppressed. 
This case will not be considered in this paper.  
The top parton density distribution in a proton is assumed to be zero,
in contrast to the other quark parton density distributions.
As a consequence, there is no gluino exchange diagram as for squark antisquark
hadroproduction.
For that reason, the $\qqbar$ channel is suppressed by a larger
power of $\beta = \sqrt{1-4 m^2_{\tilde{t}}/s}$ due to $P$-wave annihilation.
The final state must be in a state with angular momentum $l=1$
(denoted as $P$) to balance the spin of the gluon. 
Therefore, the case of stop-antistop hadroproduction needs a special treatment.  
At NLO, there is the $gq$ channel as an additional production mechanism.
At the LHC with a center of mass energy  of $7\TeV$, 
one can expect for a luminosity of $1\fb^{-1}$, $100$ to $10^4$ events;
even $10^5$ events are possible, if the stop is sufficiently light.
At the LHC with a final center of mass energy of $14\TeV$,
even more events are expected to be collected.
Hence it is necessary to predict the hadronic cross section with high
accuracy.

So far, stop pairs have been searched for at the CDF
~\cite{Aaltonen:2007sw,Ivanov:2008st,Aaltonen:2009sf,Aaltonen:2010uf} 
and D0 experiments~\cite{Abazov:2009ps,Abazov:2008kz}
at the Tevatron using different strategies, for details 
see Tab.~\ref{tab:stopsuche}.

Squarks carry colour charge, so it is not surprising that
processes involving Quantum Chromodynamics (QCD) obtain large
higher-order corrections. For the production of colour-charged
superymmetric particles, the NLO corrections have been calculated in
Ref.~\cite{Beenakker:1996ch}, NLL and approximated NNLO corrections
can be found in the 
Refs~\cite{Kulesza:2008jb,Langenfeld:2009eg,Beenakker:2009ha,Beneke:2010gm}.
It has been found that these corrections are quite sizeable.  
Electroweak NLO corrections to stop-antistop production are discussed
in Ref.~\cite{Hollik:2007wf,Beccaria:2008mi}.

The theoretical aspects of \stoppair - production up to NLO have been discussed in 
Ref.~\cite{Beenakker:1997ut} and of its NLL contributions in 
Ref.~\cite{Beenakker:2010nq}.
The hadronic LO and NLO cross section can be evaluated numerically using 
the programme \texttt{Prospino}~\cite{Beenakker:1996ed}.

In this paper, I calculate and study soft gluon effects to 
hadronic stop-antistop production in the framework of the $R$-parity-
conserving MSSM.
I use Sudakov resummation to generate the approximated
NNLO corrections and include approximated two-loop Coulomb corrections
and the exact scale dependence. 
I follow the approach for top-antitop production at the LHC
and the Tevatron~\cite{Moch:2008qy,Langenfeld:2009wd}. 

This paper is organised as follows. I review the LO and NLO
contributions to the cross section. 
Then I describe the necessary steps to construct the approximated
NNLO corrections.
Using these results I calculate the approximated NNLO cross section
and discuss the theoretical uncertainty due to scale variation
and the error due to the parton density functions (PDFs). 
I give an example how these NNLO contributions reduce the scale uncertainty
and improve exclusion limits.
 
\begin{table}[tbh]
\centering
 \begin{tabular}{llcl}
\toprule
Ref. &Process & Exclusion limit & Assumptions or comments\\
\midrule
\cite{Aaltonen:2007sw} &$\tilde{t}_1\to c\tilde{\chi}^0_1$& $\mstop[1] < 100\GeV$&
  $m_{\chi^{0}_1} > 50\GeV$\\[2mm]
\cite{Abazov:2009ps}&$p\bar{p}\to \tilde{t}_1\tilde{t}_1^\ast$ &
$130\GeV < \mstop[1] < 190\GeV$ &Comparison of theor. predictions\\[0mm]
&&&with experimental and observed limits\\[2mm]
\cite{Ivanov:2008st,Aaltonen:2009sf}&
$\tilde{t}_1\to b\tilde{\chi}^\pm_1 \to b \tilde{\chi}^0_1\ell^\pm\nu_\ell$&
$128\GeV < \mstop[1] < 135\GeV$& \\[2mm]
\cite{Abazov:2008kz,Aaltonen:2010uf}&$\tilde{t}_1\to b \ell^+ \tilde{\nu}_\ell$&
$\mstop[1] > 180\GeV$& $m_{\tilde{\nu}} \geq 45\GeV$\\
&&
$\mstop[1] = 100\GeV$& $75\GeV \leq m_{\tilde{\nu}} \leq 95\GeV$\\
\bottomrule
\end{tabular}
\caption{\small{Exclusion limits for stop searches at the Tevatron.}}
\label{tab:stopsuche}
\end{table}

\section{Theoretical Setup}
I focus on the inclusive hadronic cross section of hadroproduction 
of stop pairs, $\sigma_{p p \rightarrow \stoppair X}$, which is a function
of the hadronic center-of-mass energy $\sqrt{s}$, the stop mass $\mstop$,
the gluino mass $\mg$, the renormalisation scale $\mu_r$
and the factorisation scale $\mu_f$.
In the standard factorisation approach of perturbative QCD, it reads
\begin{eqnarray}
  \label{eq:totalcrs}
  \sigma_{pp/p\bar{p} \to \tilde{t}\tilde{t}^\ast X}(s,\mstop^2,\mg^2,\mufs,\murs) &=&
  \sum\limits_{i,j = q,{\bar{q}},g} \,\,\,
  \int\limits_{4\mstop^2}^{s }\,
  d {\shat} \,\, L_{ij}(\shat, s, \mufs)\,\,
  \hat{\sigma}_{ij \to \tilde{t}\tilde{t}^\ast} ({\shat},\mstop^2,\mg^2,\mufs,\murs)\, 
\end{eqnarray}
where the parton luminosities $L_{ij}$ are given as convolutions
of the PDFs $f_{i/p}$ defined through
\begin{eqnarray}
  \label{eq:partonlumi}
  L_{ij}(\shat, s, \mufs) &=&
  {\frac{1}{s}} \int\limits_{\shat}^s
  {\frac{dz}{z}} f_{i/p}\left(\mufs,{\frac{z}{s}}\right)
  f_{j/p}\left(\mufs,{\frac{\shat}{z}}\right)\, \enspace .
\end{eqnarray}
Here, $\shat$ denotes the partonic center of mass energy
and $\mufs, \murs$ are the factorisation and the 
renormalisation scale.
The partonic cross section is expressed by dimensionless
scaling function $f^{(kl)}_{ij}$
\begin{eqnarray}
\hat{\sigma}_{ij} &=& \frac{\as^2}{\mstop^2}
\biggl[
f^{(00)}_{ij} + 4\pi\as \Bigl(f^{(10)}_{ij} + f^{(11)}_{ij}L_N\Bigr)
+   (4\pi\as)^2\Bigl(f^{(20)}_{ij} + f^{(21)}_{ij}L_N + f^{(22)}_{ij}L_N^2 \Bigr)
\biggr]
\end{eqnarray}
with $L_N = \Ln$.
The LO scaling functions are given by~\cite{Beenakker:1997ut} 
\begin{eqnarray}
f^{(00)}_{\qqbar} &=& \frac{\pi}{54}\beta^3\rho 
\enspace= \enspace\frac{\pi}{54}\beta^3 + \mathcal{O}(\beta^5),\\
f^{(00)}_{gg} &=& \frac{\pi}{384}\rho
\biggl[41\beta -31\beta^3+\Bigl(17-18\*\beta^2+\beta^4\Bigr)
\log\biggl(\frac{1-\beta}{1+\beta}\biggr)\biggr] 
\enspace= \enspace\frac{7\pi}{384}\beta + \mathcal{O}(\beta^3)\enspace .
\end{eqnarray}
Formulas for the higher orders of the $gg$-channel 
and its threshold expansions can be found in
Refs~\cite{Beenakker:1997ut,Langenfeld:2009eg,Langenfeld:2009eu},
if one takes into account that in the case of stop-antistop
production no sum over flavours and helicities is needed.
$f^{(10)}_{gg}$ has been calculated numerically using \texttt{Prospino}
~\cite{Beenakker:1996ed}.
A fit to this function for an easier numerical handling
can be found in~\cite{Langenfeld:2009eg}. 
At NLO, $f^{(10)}_{\qqbar}$ is given at threshold by 
~\cite{Beenakker:1997ut,Beenakker:2010nq}
\begin{eqnarray}
f^{(10)}_{\qqbar} &=& \frac{f^{(00)}_{\qqbar}}{4\pi^2}
\biggl(\frac{8}{3}\*\log^2\big(8\*\beta^2\big) 
- \frac{155}{9}\*\log\big(8\*\beta^2\big) 
- \frac{\pi^2}{12\beta} + 54\*\pi\* a_{1}^{\qqbar}\biggr)\enspace.
\end{eqnarray}
The constant $a_{1}^{\qqbar}$ can be determined from a fit and is approximately
given as $a_{1}^{\qqbar} \approx 0.042\pm 0.001$.
It depends mildly on the squark and gluino masses and on the
stop mixing angle.

The $gq$-channel is absent at tree level. Its NLO contribution has been 
extracted from \texttt{Prospino}. This channel is strongly suppressed
at threshold.

The \lnbeta~terms which appear in the threshold expansions 
of the NLO scaling functions can be resummed 
systematically to all orders in perturbation theory using
the techniques described 
in~\cite{Contopanagos:1996nh,Catani:1996yz,Kidonakis:1997gm,Moch:2005ba,Czakon:2009zw}.
Logarithmically enhanced terms for the hadronic production of heavy quarks
admitting an $S$-wave are also studied in Ref.~\cite{Beneke:2009ye}
for arbitrary $SU(3)_{\text{colour}}$ representations .
The resummation is performed in Mellin space after introducing
moments $N$ with respect to the variable $\rho = 4\mstop^2/\shat$
of the physical space:
\begin{eqnarray}
  \label{eq:mellindef}
  \hat{\sigma}(N,\mstop^2) &=&
  \int\limits_{0}^{1}\,d\rho\, \rho^{N-1}\,
  \hat{\sigma}(\shat,\mstop^2)\, .
\end{eqnarray}

The resummed cross section is obtained for the individual color structures
denoted as ${\bf{I}}$ from the exponential
\begin{eqnarray}
  \label{eq:sigmaNres}
  \frac{\hat{\sigma}_{ij,\, {\bf I}}(N,\mstop^2)}
       { \hat{\sigma}^{B}_{ij,\, {\bf I}}(N,\mstop^2)}
  &=&
  g^0_{ij,\, {\bf I}}(\mstop^2) \cdot \exp\, \Big[ G_{ij,\,{\bf I}}(N+1) \Big] +
  {\cal O}(N^{-1}\log^n N) \, ,
\end{eqnarray}
where all dependence on the renormalisation and
factorisation scale $\mur$ and $\muf$ is suppressed and
the respective Born term is denoted $\hat{\sigma}^{B}_{ij,\, {\bf I}}$.
The exponent $G_{ij,\, {\bf I}}$ contains all large Sudakov logarithms $\log^k N$
and the resummed cross section~(\ref{eq:sigmaNres}) is
accurate up to terms which vanish as a power for large Mellin-$N$.
To NNLL accuracy, $G_{ij,\, {\bf I}}$ is commonly written as
\begin{eqnarray}
  \label{eq:GNexp}
  G_{ij,\, {\bf I}}(N) =
  \log N \cdot g^1_{ij}(\lambda)  +  g^2_{ij,\, {\bf I}}(\lambda)  +
  \frac{\alpha_s }{ 4 \pi}\, g^3_{ij,\, {\bf I}}(\lambda)  + \dots\, ,
\end{eqnarray}
where $\lambda = \beta_0\, \log N\, \alpha_s/(4 \pi)$.
The exponential $\exp\, \big[ G_{ij,\,{\bf I}}(N+1) \big]$ in Eq. (\ref{eq:GNexp})
is independent from the Born cross section~\cite{Moch:2008qy,Czakon:2009zw}. 
The functions $g^k_{ij}$, $k=1,2,3$, for the octet color 
structure are explicitly given
in Ref.~\cite{Moch:2008qy} and can be taken over from the case of top-quark
hadroproduction, the function $g^0_{\qqbar}$ is given by 
Eq.~(\ref{eq:g0qq}) in the App.~\ref{subsec:resum}.
All $g^k_{ij}$, $0\leq k \leq 3$, depend on a number of
anomalous dimensions, \textit{i.e.}~the well-known cusp anomalous dimension $A_q$,
the functions $D_{Q\bar{Q}}$ and $D_{q}$ controlling soft gluon emission,
and the coefficients of the QCD $\beta$-function.
The strength of soft gluon emission is proportional to the Casimir operator
of the $SU(3)_{\mathrm{colour}}$ representation of the produced state.
This is identical for $t\bar{t}$ and \stoppair - production.
Expressions for $A_q$ and $D_{q}$ are given in the Refs.~\cite{Moch:2004pa,Kodaira:1981nh},
and for $D_{Q\bar{Q}}$ in Ref.~\cite{Beneke:2009rj}. 
At higher orders, they also depend on the chosen 
renormalisation scheme, thus on the dynamical degrees of freedom.

For my fixed order NNLO calculation, I extracted the $\as^2$-terms
from the right hand side of Eq.~(\ref{eq:sigmaNres}).
At the end, I used Eqs~(\ref{eq:inveins}) - (\ref{eq:invczwei}) 
given in App.~\ref{subsec:mellininversion} to convert the Mellin space result
back to the physical $\rho$ space.
I kept all those terms which are of the order $\beta^3\lnbeta[k]$, $0\leq k \leq 4$.
Eventually, I end up with the following threshold expansion for 
$f^{(20)}_{q\bar{q}}$:   
\begin{eqnarray}
\label{eq:fqq20-num}
  f^{(20)}_{q\bar{q}} &=&
  \frac{f^{(00)}_{q\bar{q}}}{(16\pi^2)^2}\*
  \Biggl [
           \frac{8192}{9} \* \lnbeta[4]
           + \biggl(-\frac{175616}{27}+\frac{16384}{3}\*\lnzwei
           +\frac{1024}{27}\nf\biggr)\lnbeta[3]\notag\\[2mm]
&&\hspace*{15mm}    
           + \biggl( \frac{525968}{27}-\frac{87808}{3}\*\lnzwei
           -\frac{4480}{9}\*\pi^2
           +\frac{512}{3}\*C^{(1)}_{\qqbar}
           +12288\*\lnzwei[2]
           \notag\\[2mm]
&&\hspace*{20mm}
           +\frac{512}{3}\*\lnzwei\,\* \nf-\frac{2080}{9}\*\nf
           -\frac{128}{9}\frac{\pi^2}{\beta}
           \biggr)\lnbeta[2]\notag\\[2mm]
&&\hspace*{15mm}
            + \biggl(
            \frac{525968}{9}\*\lnzwei-43904\*\lnzwei[2]
            +12288\*\lnzwei[3]-\frac{4960}{9}\*C^{(1)}_{\qqbar}
            -\frac{2980288}{81}\notag\\[2mm]
&&\hspace*{20mm}
            +\frac{61376}{9}\*\zeta_3+\frac{49280}{27}\*\pi^2
            -\frac{4480}{3}\*\pi^2\*\lnzwei
		  +512\,\*C^{(1)}_{\qqbar}\*\lnzwei - 2D^{(2)}_{Q\bar{Q}}\notag\\[2mm]
&&\hspace*{20mm}
          +\Big(-\frac{128}{9}\*\pi^2+256\*\lnzwei[2]+\frac{45568}{81}
          -\frac{2080}{3}\*\lnzwei\Big)\*\nf\notag\\[2mm]
&&\hspace*{20mm}
           +\Big(\frac{266}{9}-\frac{128}{9}\*\lnzwei-\frac{4}{9}\nf \Big)
           \frac{\pi^2}{\beta}
          \biggr)\lnbeta\notag\\[2mm]
&&\hspace*{15mm}
          +\biggl(
          -\frac{13}{3} + \frac{22}{3}\lnzwei 
          +\Big(\frac{10}{27} - \frac{4}{9}\*\lnzwei\Big)\*\nf
           \biggr)\frac{\pi^2}{\beta}
+\frac{1}{27}\frac{\pi^4}{\beta^2}
+ C^{(2)}_{q{\bar q}}
    \Biggr ]\enspace .
\end{eqnarray}
$C^{(1)}_{\qqbar}$ is given as 
$C^{(1)}_{\qqbar}= 216\,\pi\, a^{\qqbar}_1 -\tfrac{310}{27}$,
$C^{(2)}_{q{\bar q}}$ is the unknown 2-loop matching constant,
which is set to zero in the numerical evaluation
and $D^{(2)}_{Q\bar{Q}} = 460 -12\pi^2+72\zeta_3-\tfrac{88}{3}\nf$, see Ref.~\cite{Beneke:2009rj}.
The $\beta^3$-behaviour of the threshold expansion of the LO cross section
comes \emph{only} from the $P$-wave of the final state $\stoppair$ as mentioned in the 
introduction and does not spoil the factorisation properties in the threshold
region of the phase space.
Note that the formulas given in Ref.~\cite{Czakon:2009zw} can easily extended
to Mellin transformed cross sections $\omega$, which vanish as a power $\beta^k$
with $k\geq 1$.
These two more powers of $\beta$ in the \qqbar-channel of \stoppair - production 
lead to an additional $1/N$ factor in Mellin space.
Eq.~(\ref{eq:sigmaNres}) reproduces the known NLO threshold expansion
given in Ref.~\cite{Beenakker:1997ut,Beenakker:2010nq} for the $q\bar{q}$
channel.
This is a check that the approach works. 
Logarithmically enhanced terms which are suppressed by an additional
$1/N$ factor appear in the resummation of the (sub)leading $\log^k(N)/N$-terms 
of the corrections for the structure function $F_L$ and
are studied in detail in the Refs~\cite{Moch:2009hr,Laenen:2008gt}.  
For these reasons I apply the formulas derived for heavy quark hadroproduction. 

The coefficients of the $\ln^4\beta$, $\ln^3\beta$, and $\ln^2\beta$ 
terms depend only on first order anomalous dimensions and on the constant
$C^{(1)}_{\qqbar}$ which is related to the NLO constant  $a^{\qqbar}_1$,
see the equation above.
The linear $\log\beta$ term depends on $C^{(1)}_{\qqbar}$ as well and on other 
first order (NLO) contributions,
but also on second-order anomalous dimensions and non-Coulomb potential 
contributions~\cite{Beneke:2009ye}.
In Tab.~\ref{tab:terms}, I show for four examples how these parts contribute
to the hadronic NNLO threshold corrections. 
The numbers show that terms which have an NLO origin contribute
most and that NNLO contributions have a small but sizeable effect.   

I also included the Coulomb corrections up to NNLO.
For the singlet case, the Coulomb contributions are studied in Ref.~\cite{Czarnecki:1997vz}.
Generalisation to other colour structures requires the substitution
of the corresponding group factors and decomposition of the 
colour structures of the considered process in irreducible colour representations.
The last step will not be necessary for stop pair production in the $q\bar{q}$-annihilation
channel.
The NLO Coulomb corrections agree with the NLO Coulomb corrections 
for top antitop production~\cite{Langenfeld:2009eg,Beenakker:1997ut}.
In both cases, only the colour octet contributes to the scaling function
at the corresponding leading order in $\beta$.
Therefore, I have used as an approximation for the NNLO Coulomb contributions 
for $\tilde{t}\tilde{t}^\ast$-production the same NNLO Coulomb contributions
as for $t\bar{t}$-production~\cite{Czarnecki:1997vz,Moch:2008qy,Langenfeld:2009eg}.
Gauge invariance together with Supersymmetry support this approximation.
Note that the $\log^2\beta/\beta$-term comes from interference of the NLO
Coulomb-contribution with the NLO threshold logarithms.
Tab.~\ref{tab:terms} shows also the NLO and the pure NNLO Coulomb contributions to the 
NNLO threshold corrections at the hadronic level. 

The scale-dependent scaling functions are derived by renormalisation group
techniques following Refs~\cite{vanNeerven:2000uj,Kidonakis:2001nj}:
\begin{eqnarray}
\label{eq:f11}
f_{ij}^{(11)}&=&
\frac{1}{16\pi^2}\*\left(
  2\* \beta_0 \*f_{ij}^{(00)} - f_{kj}^{(00)}\otimes P_{ki}^{(0)} 
- f_{ik}^{(00)}\otimes P_{kj}^{(0)}
\right)
\, ,\\[2mm]
\label{eq:f21}
f_{ij}^{(21)}&=&
\frac{1}{(16\pi^2)^2}\*\left(
  2\* \beta_1\* f_{ij}^{(00)} - f_{kj}^{(00)}\otimes P_{ki}^{(1)}
  - f_{ik}^{(00)}\otimes P_{kj}^{(1)}\right)
\nonumber\\
& &
+ \frac{1}{16\pi^2}\*\left(
  3 \*\beta_0 \*f_{ij}^{(10)}
  -f_{kj}^{(10)}\otimes P_{ki}^{(0)}
  - f_{ik}^{(10)}\otimes
  P_{kj}^{(0)}\right)
\, ,
\\[2mm]
\label{eq:f22}
f_{ij}^{(22)}&=&
\frac{1}{(16\pi^2)^2}\*\left(
  f_{kl}^{(00)}\otimes P_{ki}^{(0)}\otimes P_{lj}^{(0)}
  +\frac{1}{2} f_{in}^{(00)}\otimes P_{nl}^{(0)}\otimes P_{lj}^{(0)}
  +\frac{1}{2} f_{nj}^{(00)}\otimes P_{nk}^{(0)}\otimes P_{ki}^{(0)}
									   \right.\notag\\[2mm]
& & \hspace{18mm}\left. + 3 \*\beta_0^2  \*f_{ij}^{(00)}
  - \frac{5}{2}\*\beta_0 \*f_{ik}^{(00)}\otimes P_{kj}^{(0)}
  - \frac{5}{2}\*\beta_0 \*f_{kj}^{(00)}\otimes P_{ki}^{(0)}
\right)
\, ,
\end{eqnarray}
where $\otimes$ denotes the standard Mellin convolution;
these are ordinary products in Mellin space using Eq.~(\ref{eq:mellindef}). 
Repeated indices imply summation over admissible partons.
However, I restrict myself for phenomenological applications  to the
numerically dominant diagonal parton channels at two-loop.
Note that the scale dependence is exact at all energies, even away from
threshold, 
because the Eqs (\ref{eq:f11})-(\ref{eq:f22}) depend on functions 
which are at least one order lower than they themselves have.  
The functions $P_{ij}(x)$ are called splitting functions and govern the PDF
evolution. 
They have the expansion
\begin{eqnarray}
\label{eq:splitting}
P_{ij}(x) &=& \frac{\as}{4\pi}P_{ij}^{(0)}(x) + 
 \left(\frac{\as}{4\pi}\right)^2P_{ij}^{(1)}(x) + \ldots .
\end{eqnarray}
Explicit expressions for the $P_{ij}^{(k)}$ can be found in
Refs~\cite{Moch:2004pa,Vogt:2004mw}.

Analytical results for $f^{(11)}_{gg}$ and $f^{(11)}_{\qqbar}$
are given in Ref.~\cite{Langenfeld:2009eu}.
For the $gq$-channel, Eq.~(\ref{eq:f11}) simplifies to
\begin{eqnarray}
\label{eq:sqg11}
f^{(11)}_{gq} &=& -\frac{1}{16\pi^2}\left(P^{(0)}_{gq}\otimes f^{(0)}_{gg}
		+\frac{1}{2 n_f}P^{(0)}_{qg}\otimes f^{(0)}_{\qqbar}\right).
\end{eqnarray} 
The integration can be done explicitly yielding
\begin{align}
f^{(11)}_{gq} \,\,=\,\,\,\,\,&\frac{1}{51840\*\pi}\*
\biggl[
\beta\*\Big(-176 - 1083\*\rho +1409\*\rho^2\Big)
+15\*\rho\*\Big(26 - (27-24\*\ln 2)\*\rho - 4\*\rho^2\Big)\*L_2 \notag\\[2mm]
&\hspace*{17mm}+180\*\rho^2\*\Big(2\*L_4-L_6\Big)
\biggr],
\end{align} 
where the functions $L_2$, $L_4$, and $L_6$~\cite{Langenfeld:2009eu} are
defined as
\begin{eqnarray}
L_2 = \log\left(\tfrac{1+\beta}{1-\beta}\right),\enspace
L_4 = \Li\left(\tfrac{1-\beta}{2}\right) - \Li\left(\tfrac{1+\beta}{2}\right),
	 \enspace
L_6 = \log^2(1-\beta) - \log^2(1+\beta).
\end{eqnarray}
The high energy limit of this scaling function is
\begin{eqnarray}
\lim_{\beta \to 1} f^{(11)}_{gq} = -\frac{11}{3240\*\pi},
\end{eqnarray}
which agrees with the result given in Ref.~\cite{Beenakker:1997ut}.

The threshold expansions of the NNLO-scale-dependent 
scaling functions of the $\qqbar$ channel read
\begin{eqnarray}
f^{(21)}_{\qqbar} &=& -\frac{f_{\qqbar}^{(00)}}{(16\pi^2)^2}
\biggl[
{\frac {8192}{9}}\, \lnbeta[3]
+ \left( {\frac {256}{3}}\,\nf+{\frac {32768}{9}}\,\lnzwei 
-{\frac {46976}{9}} \right)  \lnbeta[2] \notag\\[2mm]
&&\hspace*{5mm}
+ \biggl( -{\frac {383104}{27}}\,\lnzwei 
+{\frac {798872}{81}}+{\frac {14336}{3}}\, \lnzwei[2]
-{\frac {8080}{27}}\,\nf-{\frac {2240}{9}}\,{\pi }^{2}
-{\frac {64}{9}}\,{\frac {{\pi }^{2}}{\beta}} \notag\\[2mm]
&&\hspace*{5mm}
+{\frac {256}{3}}\,C^{(1)}
+256\,\nf\,\lnzwei  \biggr) \lnbeta  
+{\frac {4540}{81}}\,\nf-{\frac {1924}{9}}\,C^{(1)}
+2048\,\lnzwei[3] \notag\\[2mm]
&&\hspace*{5mm}
+{\frac {393004}{27}}\,\lnzwei -{\frac {1449488}{243}}
+8\,C^{(1)}\,\nf-{\frac {85856}{9}}\, \lnzwei[2]
+{\frac {14240}{9}}\,\zeta_3  \notag\\[2mm]
&&\hspace*{5mm}
+{\frac {11024}{27}}\,{\pi }^{2}-{\frac {1088}{3}}\,{\pi }^{2}\lnzwei 
+192\, \lnzwei[2]\,\nf
+{\frac {25}{3}}\,{\frac {{\pi }^{2}}{\beta}}
+{\frac {256}{3}}\,C^{(1)}\,\lnzwei  \notag\\[2mm]
&&\hspace*{5mm}
-{\frac {11800}{27}}\,\nf\,\lnzwei 
-{\frac {32}{27}}\,{\pi }^{2}\nf-\frac{2}{3}\,\nf {\frac {{\pi }^{2}}{\beta}}
\biggr],\\[3mm]
f^{(22)}_{\qqbar} &=& \frac{f_{\qqbar}^{(00)}}{(16\pi^2)^2}
\biggl[
{\frac {2048}{9}}\, \lnbeta[2]+
 \left( -{\frac {27616}{27}}+{\frac {320}{9}}\,\nf
+{\frac {4096}{9}}\,\lnzwei  \right) \lnbeta  \notag\\[2mm]
&&\hspace*{5mm}
-{\frac {2108}{27}}\,\nf+{\frac {112351}{81}}
-{\frac {27616}{27}}\,\lnzwei
+{\frac {2048}{9}}\, \lnzwei[2]
+{\frac {320}{9}}\,\nf\,\lnzwei  \notag\\[2mm]
&&\hspace*{5mm}
-{\frac {256}{9}}\,{\pi }^{2}+\frac{4}{3}\,n_f^{2}
\biggr]
\end{eqnarray}
with $C^{(1)} = 54\,\pi\,a_1^{\qqbar}$. 

\begin{figure}
  \centering
  \scalebox{0.37}{\includegraphics{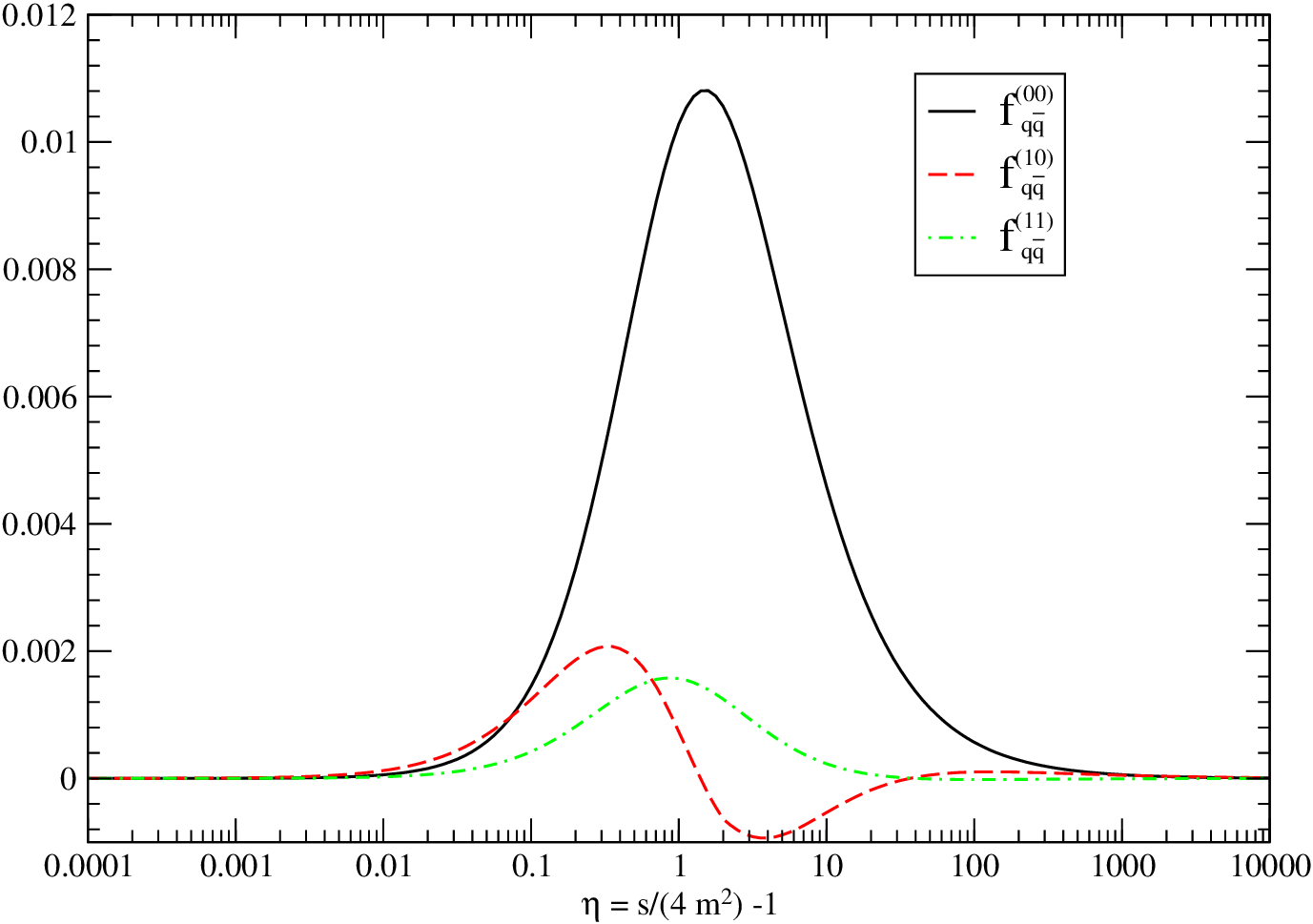}}
  \vspace*{10mm}

  \scalebox{0.37}{\includegraphics{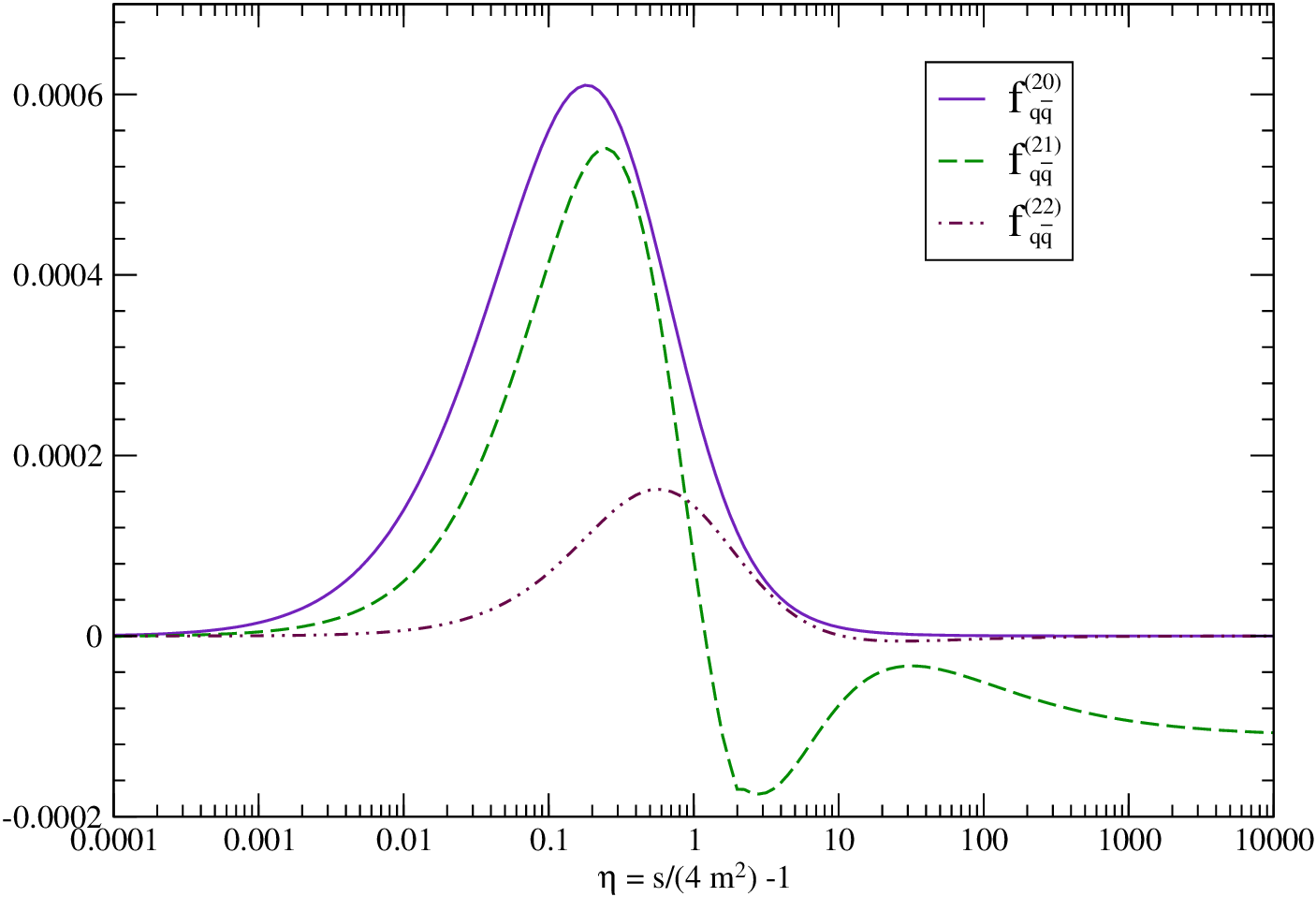}}
  \vspace*{10mm}
  
  \scalebox{0.37}{\includegraphics{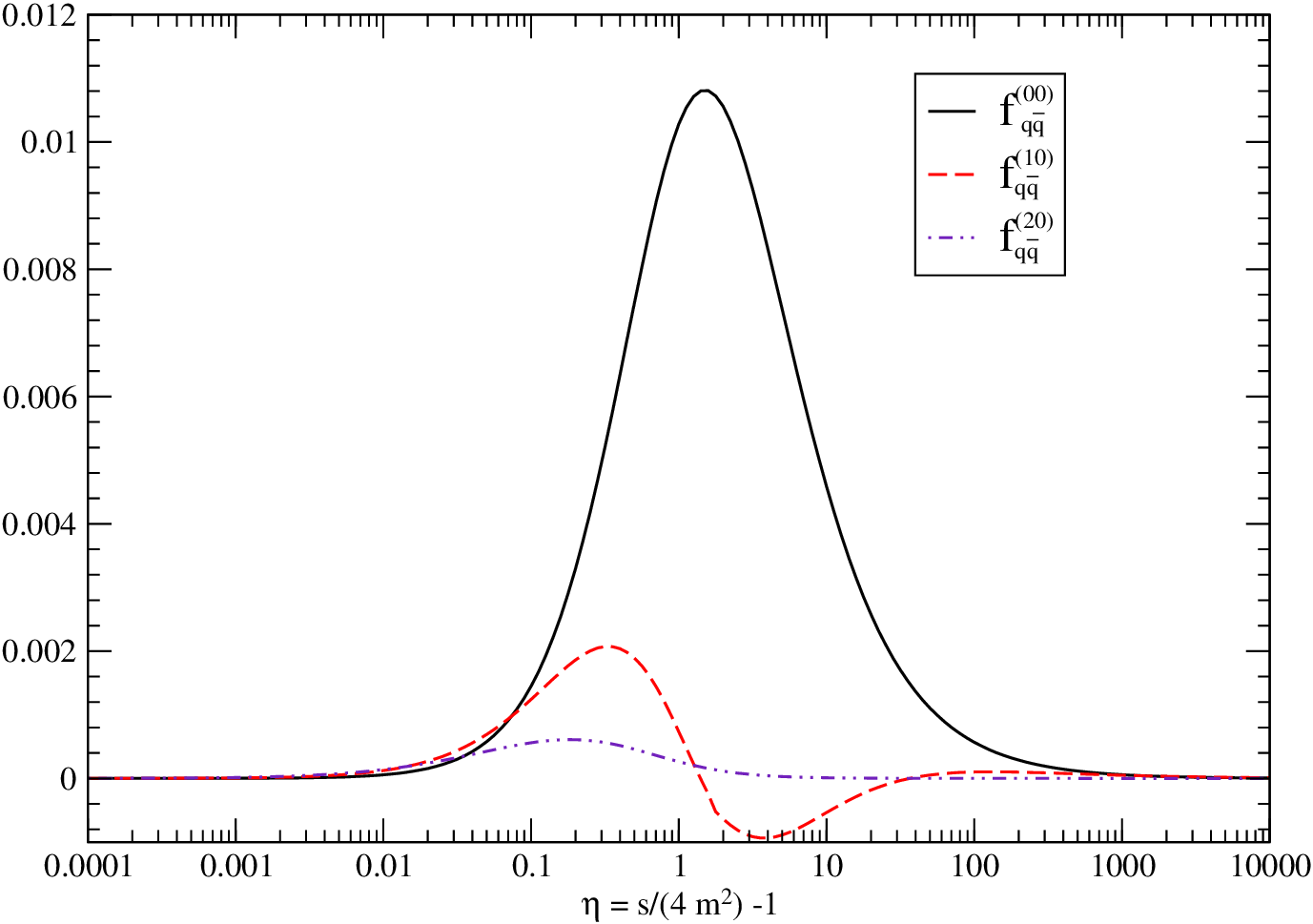}}
  \vspace*{2mm}
  \caption{\small{Scaling functions $f^{(ij)}_{\qqbar}$ with $i = 0,1,2$ and $j\le i$.
    The masses are  $\mstop[1]=300\,\GeV$, 
			  $\mq = 400\,\GeV$, 
			  $\mstop[2]=480\,\GeV$
			  $\mg = 500\,\GeV$.}}
  \label{fig:scalingfunctions}
\end{figure}

In Fig.~\ref{fig:scalingfunctions}, I show the LO, NLO, and NNLO
scaling functions.
The scaling functions $f^{(00)}_{\qqbar}$, $f^{(11)}_{\qqbar}$
$f^{(20)}_{\qqbar}$, and $f^{(22)}_{\qqbar}$
depend only on the dimensionless variable 
$\eta = \tfrac{\hat{s}}{4\mstop^2}-1$, 
but $f^{(10)}_{\qqbar}$ and $f^{(21)}_{\qqbar}$ 
depend also mildly on the masses of the squarks and the gluino
and the stop mixing angle~\cite{Beenakker:1997ut}.
At the hadronic level, the effect for the NLO + NLL cross section is 
smaller than $2\%$~\cite{Beenakker:2010nq}, so I neglect them.    

As example point, I have chosen the following masses:
$\mstop[1]=300\,\GeV$,
$\mq = 400\,\GeV = 1.33\,\mstop[1]$,
$\mstop[2]=480\,\GeV = 1.6\,\mstop[1]$,
$\mg = 500\,\GeV = 1.67\,\mstop[1]$,
and $\theta =\pi/2$, \textit{i.e.}~$\mstop[1]=\mstop[\text{R}]$ and
$\mstop[2]=\mstop[\text{L}]$.
When varying the stop mass I conserve these mass relations.
I restrict myself to the lighter stop, but the results
also apply to the heavier stop, because the gluon-stop-stop
interactions entering my process do not distinguish between
the left-handed and the right-handed stop squarks.
 
\begin{table}
\centering
\begin{tabular}{cc r rrr r rrr r}
\toprule[0.08em]
 Collider&$\mstop$&$ \sum$\phantom{N} & $\ln^4(\beta)$ & $\ln^3(\beta)$ &$\ln^2(\beta)$ & 
  \multicolumn{3}{c}{$\log(\beta)$} &C$_{\rm{NLO}}$&C$_{\rm{NNLO}}$\\
 &&&&&&nC&$D^{(2)}_{Q\bar{Q}}$&rest&&\\
\midrule[0.03em]
LHC $14 \TeV$&$300\GeV$&74.54&5.33&18.80&31.87&\multicolumn{3}{c}{27.71}&-9.45& 0.29\\[1mm]
 &&&&&&-2.23&17.90&12.04&&\\[1mm]
\hline
LHC $14 \TeV$&$600\GeV$& 2.93&0.24& 0.81& 1.28&\multicolumn{3}{c}{0.97}& -0.37&-0.01\\[1mm]
 &&&&&&-0.08&0.63&0.42&&\\[1mm]
\hline
LHC $ 7 \TeV$&$300\GeV$&17.5&1.42& 4.83& 7.66&\multicolumn{3}{c}{5.85}& -2.21&-0.06\\[1mm]
 &&&&&&-0.47&3,78&2.54&&\\[1mm]
\hline
Tevatron     &$300\GeV$&1.41 &0.16& 0.48& 0.62&\multicolumn{3}{c}{0.34}& -0.17&-0.02\\
&&&&&&-0.03&0.22&0.15&&\\[1mm]
\bottomrule[0.08em] 
\end{tabular}
\caption{\small{Individual hadronic contributions of the $\log$-powers 
and Coulomb-corrections to the NNLO threshold contributions of the
\qqbar-channel in $\fb$. 
$\ln^4(\beta)$ has to be understood as 
$\tfrac{f^{(00)}_{\qqbar}}{(16\pi^2)^2}\cdot \tfrac{8192}{9}\ln^4(\beta)$, 
analogously for the other terms. 
The linear $\log$-term is decomposed into contributions coming from
non-Coulomb potential terms, from the two loop anomalous dimension $D^{(2)}_{Q\bar{Q}}$,
and, finally,  the rest.
The Coulomb contribution are decomposed into contributions coming from the interference
of NLO threshold logarithms with NLO Coulomb corrections C$_{\rm{NLO}}$
and pure NNLO Coulomb corrections C$_{\rm{NNLO}}$. 
$\sum$ denotes the sum over all NNLO threshold contributions.
The PDF set used is MSTW 2008 NNLO~\cite{Martin:2009iq}.}}
\label{tab:terms}
\end{table}

\section{Results}
\subsection{Hadronic cross section}
I start with the discussion of the total hadronic cross section,
which is obtained by convoluting the 
partonic cross section with the PDFs, see Eq.~(\ref{eq:totalcrs}).
I keep the $gg$ and the $\qqbar$ channel at all orders up to NNLO,
and for the scale-dependent terms, only contributions coming from diagonal parton 
channels are considered.
Only the NLO contributions of the $gq$ channel are considered, 
which are the leading contributions of this channel.
The scale dependence of this channel is given by Eq.~(\ref{eq:sqg11}). 
 
I define the NLO and NNLO $K$ factors as
\begin{eqnarray}
\label{eq:kfactor}
K_{\text{NLO}} = \frac{\sigma^{\text{NLO}}}{\sigma^{\text{LO}}},\quad
K_{\text{NNLO}} = \frac{\sigma^{\text{NNLO}}}{\sigma^{\text{NLO}}}\enspace .
\end{eqnarray}
I use for all orders MSTW2008 NNLO PDFs~\cite{Martin:2009iq},
if not otherwise stated.
Therefore, the $K$ factors account only for the pure higher order 
corrections of the partonic cross section (convoluted with the PDFs)
and not for higher order
corrections of the PDFs and the strong coupling constant $\as$. 
 
\begin{figure}[t!]
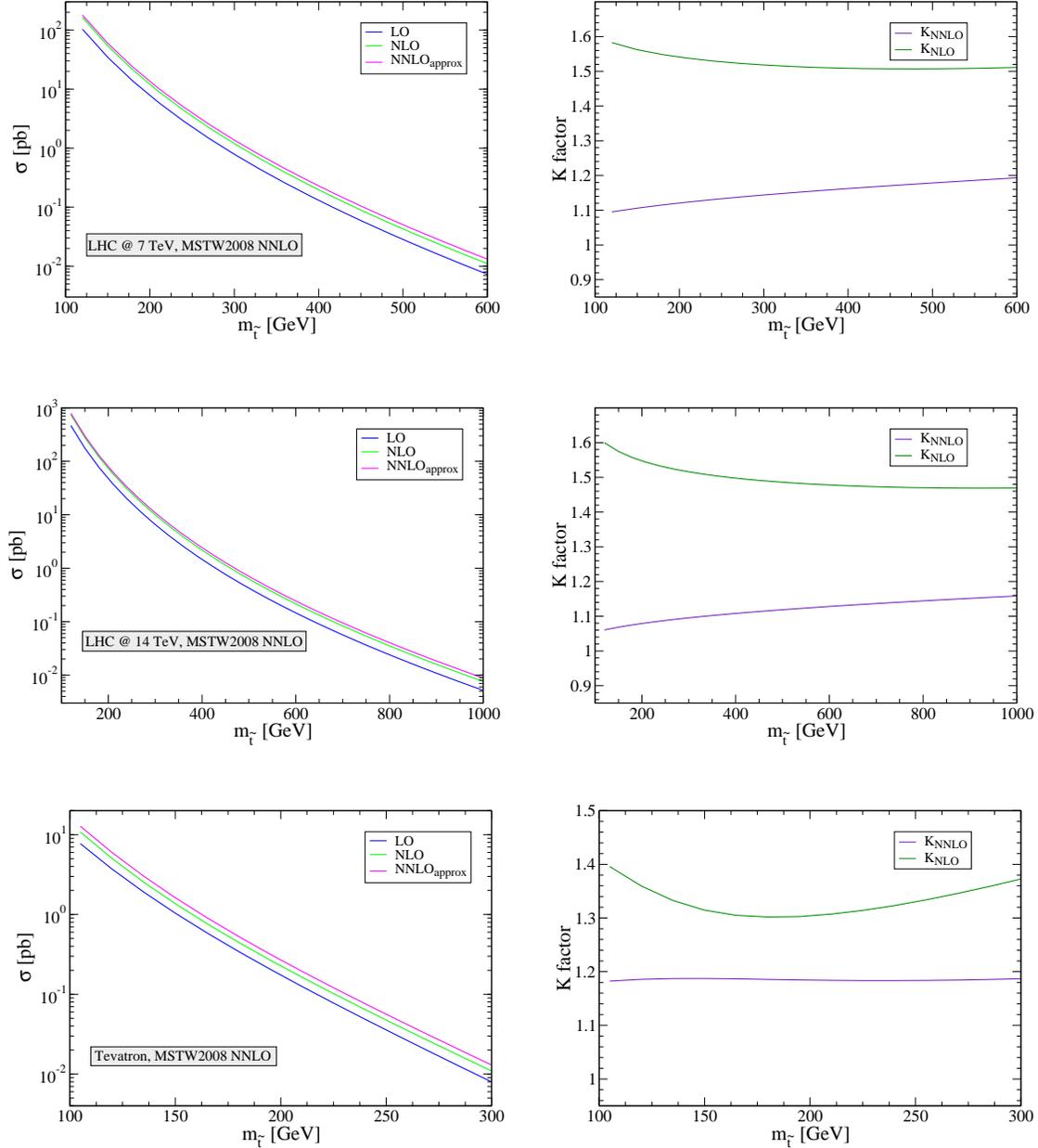

\centering
\vspace*{15mm}
\scalebox{0.28}{\includegraphics{./pics/lhc07mstw2008totalxsecA.eps}}
\hspace*{5mm}
\scalebox{0.28}{\includegraphics{./pics/lhc07mstw2008totalxsecB.eps}}
\vspace*{8mm}

\scalebox{0.28}{\includegraphics{./pics/lhc14mstw2008totalxsecA.eps}}
\hspace*{5mm}
\scalebox{0.28}{\includegraphics{./pics/lhc14mstw2008totalxsecB.eps}}
\vspace*{8mm}

\scalebox{0.28}{\includegraphics{./pics/tevamstw2008totalxsecA.eps}}
\hspace*{5mm}
\scalebox{0.28}{\includegraphics{./pics/tevamstw2008totalxsecB.eps}}    
\caption{\small{Total hadronic cross section at LO, NLO, and  NNLO$_{\text{approx}}$
     at the LHC 7\TeV (first row) and 14\TeV (second row) and
	the Tevatron (1.96\TeV, third row).
     The right column shows the corresponding $K$ factors. 
     The PDF set used is MSTW2008 NNLO~\cite{Martin:2009iq}.}}
  \label{fig:tothadxsec14}
\end{figure}

In the left column of Fig.~\ref{fig:tothadxsec14}, 
I show the total hadronic cross section
for the LHC ($7\TeV$ first row, $14\TeV$ second row) and the Tevatron 
(third row) as a function of the stop mass. 
Similar to  top-antitop~\cite{Moch:2008qy,Langenfeld:2009wd} 
and squark-antisquark production~\cite{Langenfeld:2009eg},
the total cross section shows a strong mass dependence. 
At the LHC ($14\TeV$), the cross section decreases 
within the shown stop mass range from about $1000\pb$ to $10^{-2}\pb$. 
In the right column of Fig.~\ref{fig:tothadxsec14}, 
I show the corresponding $K$ factors.

For example, for a stop mass of $300\GeV$ produced at the LHC with $14\TeV$, 
I have a total cross section
of $6.57\pb$, $9.96\pb$, $10.92\pb$ at LO, NLO, NNLO$_\text{approx}$, 
respectively.
For a stop mass of $600\GeV$, I find $0.146\pb$, $0.216\pb$, $0.244\pb$.
The $K$-factors are $K_{\text{NLO}}\approx 1.5$  and 
$K_{\text{NNLO}}\approx 1.1$ and $1.13$.
For the LHC at $7\TeV$, one finds similar values for stop masses in the 
interval $100\GeV \leq \mstop \leq 600\GeV$.
At the Tevatron, I have $K_{\text{NLO}} = 1.3\ldots 1.4$
and $K_{\text{NNLO}}\approx 1.2$ for stop masses in the range
$100\GeV \leq \mstop \leq 300\GeV$.

In Tabs~\ref{tab:xsecvalues} - \ref{tab:xsecvaluesteva}, 
see App.~\ref{subsec:tables}, values for the total hadronic cross section
for different masses, PDF sets, scales and colliders are shown.
The values for the PDF sets Cteq6.6~\cite{Nadolsky:2008zw},
MSTW 2008 NNLO~\cite{Martin:2009iq},and CT10~\cite{Lai:2010vv} show only 
small differences, whereas the ABKM09 NNLO (5 flavours) PDFs differ in
the treatment of the gluon PDF from the other PDF sets. 
This leads to sizeable differences in the total cross sections.

\subsection{Theoretical and Systematic Uncertainty and PDF error}
\begin{figure}
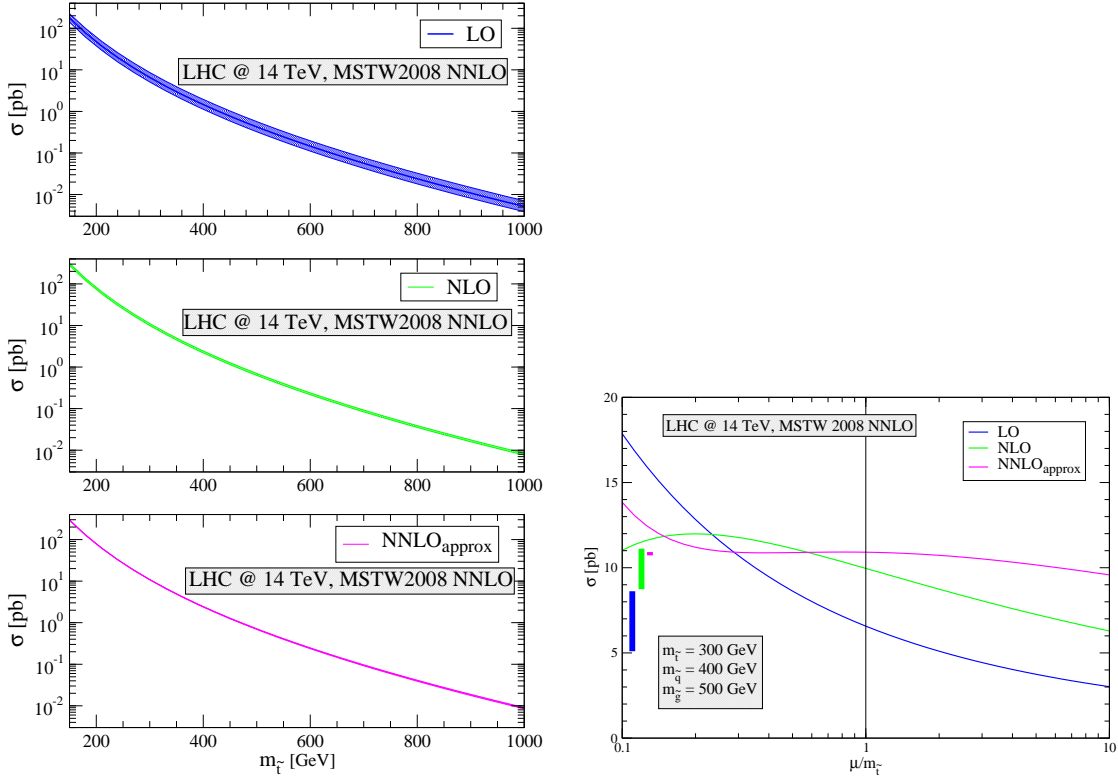

  \centering
  \scalebox{0.4}{\includegraphics{./pics/lhc14mstwcompscaleuncert.eps}}
  \hspace*{3mm}
  \scalebox{0.3}{\includegraphics{./pics/scalevarlhc14.eps}}
  \caption{\small{Left hand side: Theoretical uncertainty of the total hadronic cross section at 
	the LHC (14\TeV) at LO (upper figure, blue band), 
	NLO (central figure, green band), 
	NNLO$_{\text{approx}}$ (lower figure, purple line). 
	At NNLO$_{\text{approx}}$, the theoretical uncertainty has shrunk
	to a small band.
     Right hand side: Scale dependence of the total hadronic cross section
     for the example point $\mstop[1] = 300\GeV$, 
					  $\mq = 400\,\GeV$, 
					  $\mstop[2]=480\,\GeV$,
					  $\mg = 500\,\GeV$.
    The vertical bars indicate the total scale variation
    in the range $[\mstop/2,2\*\mstop]$.} 
   }
  \label{fig:tothadtheorunc}
\end{figure}

In this section, I address the following sources for errors:
the systematic theoretical error, the scale uncertainty,
and the PDF error.

In Tab.~\ref{tab:terms}, I listed the individual $\lnbeta[k]$ contributions
to the total NNLO contributions. Note that the NNLO matching constants 
$C^{(2)}_{ij}$ are unknown and set to zero. Compared to the total NNLO
contributions the $\lnbeta[1]$ term is quite sizeable, this translates
to a roughly $3-5\%$ contribution to the NNLO cross section.
To estimate the systematic error coming from the NNLO matching constants 
$C^{(2)}_{ij}$, I proceed as described in Ref.~\cite{Langenfeld:2009wd}.
I find for the ratio $\sigma_{\text{NLL+Coul}}/\sigma_{\text{exact}}
= 1.10\ldots 1.25$. 
This ratio translates to an estimate for the relative systematic error coming
from the  NNLO matching constants as $1-2.5\%$.   

The total hadronic LO, NLO, and NNLO cross sections are shown on
the left of Fig.~\ref{fig:tothadtheorunc}  
as a function of the stop mass 
and for variations of the scale  $\mu$ with 
$\mstop/2 \leq \mu \leq  2\mstop$, where
I have identified the factorisation scale with the renormalisation scale.  
The width of the band indicates the scale uncertainty, which
becomes smaller when going from LO to NLO and NNLO.
On the right-hand side of Fig.~\ref{fig:tothadtheorunc},
the scale dependence for the example point is shown in more detail. 
I quote as theoretical uncertainty 
\begin{eqnarray}
 \min\sigma(\mu) \leq \sigma(\mstop) \leq \max\sigma(\mu),
\end{eqnarray}
where the $\min$ and $\max$ are to be taken over the interval
$[\mstop[1]/2, 2\mstop[1]]$.
At LO and NLO, the minimal value is taken at $\mu=2\,\mstop$
and the maximal value at  $\mu = \mstop/2$.
However, this is not longer true at NNLO. 
For the theoretical error, one finds
\begin{eqnarray}
\sigma_{\text{LO}} = 6.57^{+2.06}_{-1.43}\pb,\quad
\sigma_{\text{NLO}} = 9.96^{+1.17}_{-1.22}\pb,\quad
\sigma_{\text{NNLO}} = 10.90^{+0.01}_{-0.18}\pb \enspace .
\end{eqnarray}
As one can see, there is a strong scale dependence at LO,
becoming weaker at NLO,
and is flattend out at NNLO within the considered range.
This flattening gives a hint that the approach is reliable.

\begin{figure}
  \centering
  \scalebox{0.7}{\includegraphics{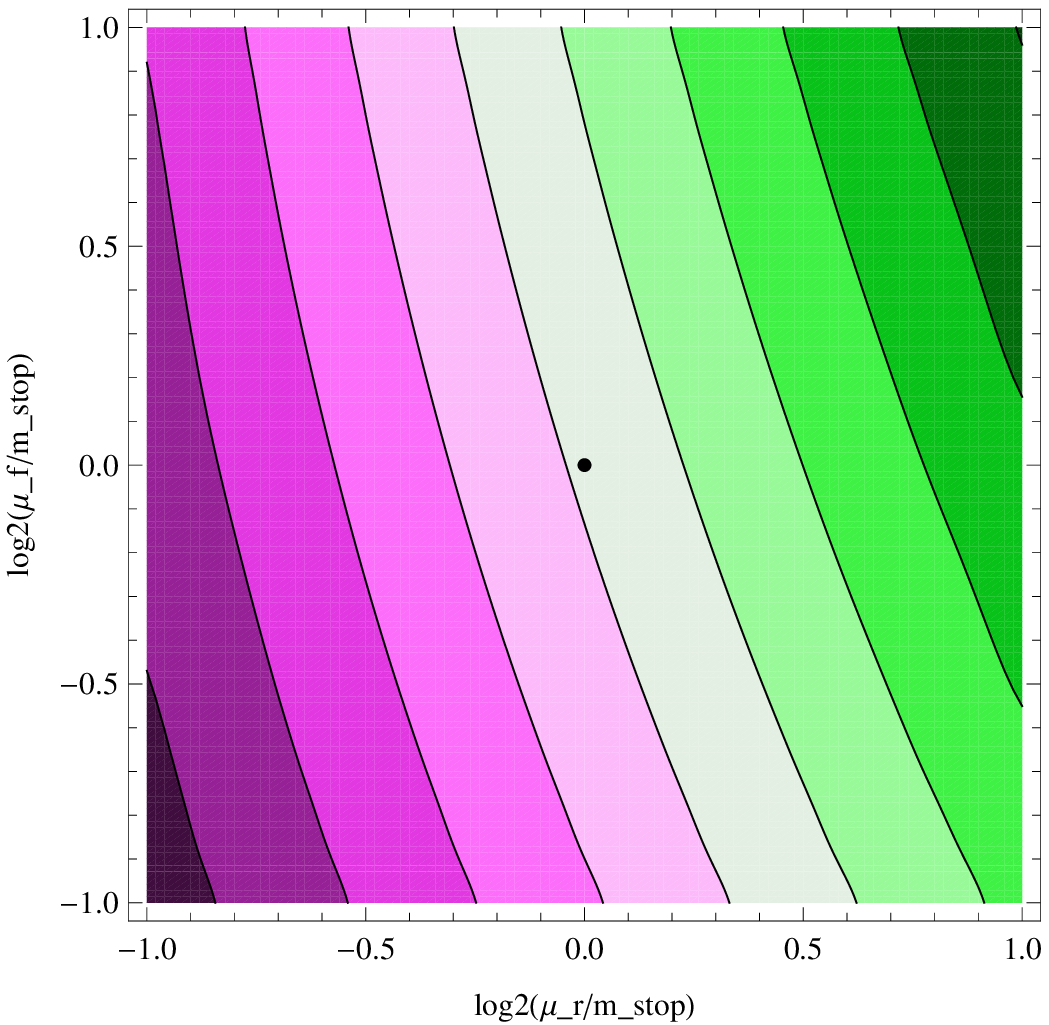}
	\put(-247,50){\rotatebox{106}{${11\pb}$}}
	\put(-220,70){\rotatebox{108}{${10.75\pb}$}}
	\put(-190,90){\rotatebox{106}{${10.5\pb}$}}
	\put(-163,110){\rotatebox{108}{${10.25\pb}$}}
	\put(-132,130){\rotatebox{108}{${10\pb}$}}
	\put(-107,150){\rotatebox{108}{${9.75\pb}$}}
	\put(-78,170){\rotatebox{108}{${9.5\pb}$}}
	\put(-52,190){\rotatebox{108}{${9.25\pb}$}}
	\put(-22,210){\rotatebox{106}{${9\pb}$}}
	}
  \hspace*{3mm}
  \scalebox{0.7}{\includegraphics{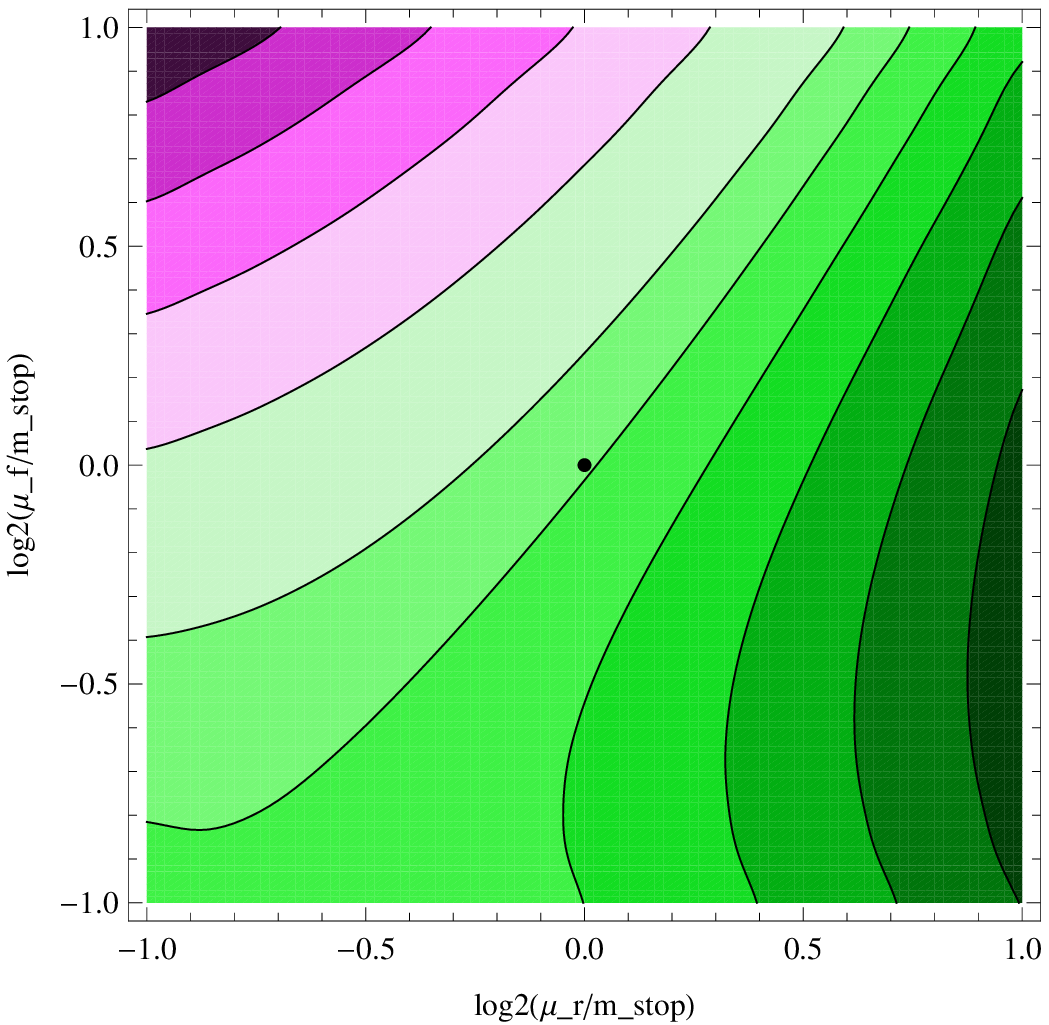}
	\put(-247,263){\rotatebox{32}{${11.8\pb}$}}
	\put(-229,243){\rotatebox{32}{${11.6\pb}$}}
	\put(-207,223){\rotatebox{35}{${11.4\pb}$}}
	\put(-174,203){\rotatebox{39}{${11.2\pb}$}}
	\put(-136,183){\rotatebox{48}{${11\pb}$}}
	\put(-122,163){\rotatebox{52}{${10.9\pb}$}}
	\put(-104,143){\rotatebox{58}{${10.8\pb}$}}
	\put(-78,123){\rotatebox{71}{${10.7\pb}$}}
	\put(-51,108){\rotatebox{79}{${10.6\pb}$}}
	\put(-20,103){\rotatebox{80}{${10.5\pb}$}}
	}
  \caption{\small{Contour lines of the total hadronic NLO (left) 
	and NNLO (right) cross section from the 
	independent variation of the renormalisation and factorisation
	scale $\mu_r$ and $\mu_f$ for LHC, $14\TeV$, 
	with PDF set MSTW2008 NNLO~\cite{Martin:2009iq}
	for the example point with $\mstop=300\GeV$.
	The dot in the middle of the figure indicates the cross section
	for $\mu_r=\mu_f=\mstop$, and the range corresponds to
	$\mu_f,\mu_r \in [\mstop/2,2\mstop]$.}
	}
  \label{fig:murmuf}
\end{figure}

Using renormalisation group techniques, one recovers the full 
dependence on the renormalisation scale $\mu_r$
and factorisation scale $\mu_f$.
I have  done this for the example point for the NLO and the
NNLO cross section, see Fig.~\ref{fig:murmuf}.
I define the theoretical uncertainty coming from an independent variation of
$\mu_r$ and $\mu_f$ in the standard range  
$\mu_r$, $\mu_f\in [\mstop/2, 2\mstop]$ as
\begin{eqnarray}
\label{eq:inderror}
\min \sigma(\mu_r,\mu_f) \leq \sigma(\mstop) \leq \max\sigma(\mu_r,\mu_f).
\end{eqnarray} 
The contour lines of the total cross section for the example point
with an independent variation
of $\mu_r$ and $\mu_f$ are shown in Fig.~\ref{fig:murmuf}.
Note that the range of the axes is from $\log_2(\mu_{r,f}/\mstop[1]) = -1$ 
to $\log_2(\mu_{r,f}/\mstop[1]) =1$.
The scale variation with fixed scales $\mu_r=\mu_f$ proceeds
along the diagonal from the lower left to the upper right corner
of the figure.
The gradient of the NLO contour lines lies approximately in the
$\mu_r=\mu_f$ direction, meaning that the theoretical error
from the definition in Eq.~(\ref{eq:inderror}) is the same
as if one sets $\mu_r=\mu_f$.
For the NNLO case one observes the opposite situation: 
the contour lines are nearly parallel to the diagonal $\mu_r=\mu_f$. 
I obtain a larger uncertainty in that case:
\begin{eqnarray}
\sigma_{\text{NLO}} = 9.96^{+1.17}_{-1.22}\pb,\quad
\sigma_{\text{NNLO}} = 10.90^{+1.05}_{-0.46}\pb.
\end{eqnarray}
 
Another source of error to discuss is the PDF error.
I calculated the PDF uncertainty according to Ref.~\cite{Nadolsky:2008zw}
for the two PDF sets CT10 and MSTW2008 NNLO (90\% C.L.).
In both cases, the uncertainty increases with higher stop masses
due to large uncertainties of the gluon PDF in high $x$-ranges.
For CT10, I find as relative errors $\approx 3\%$ for $\mstop=100\GeV$ 
and $\approx 18\%$ for $\mstop=1000\GeV$,
and for MSTW2008 NNLO, the relative errors are $\approx 3\%$ for 
$\mstop=100\GeV$ and  $\approx 10\%$ for $\mstop=1000\GeV$.
The relative error of the MSTW2008 NNLO PDF set is smaller for large
stop masses compared to the CT10 PDFs.

\begin{figure}
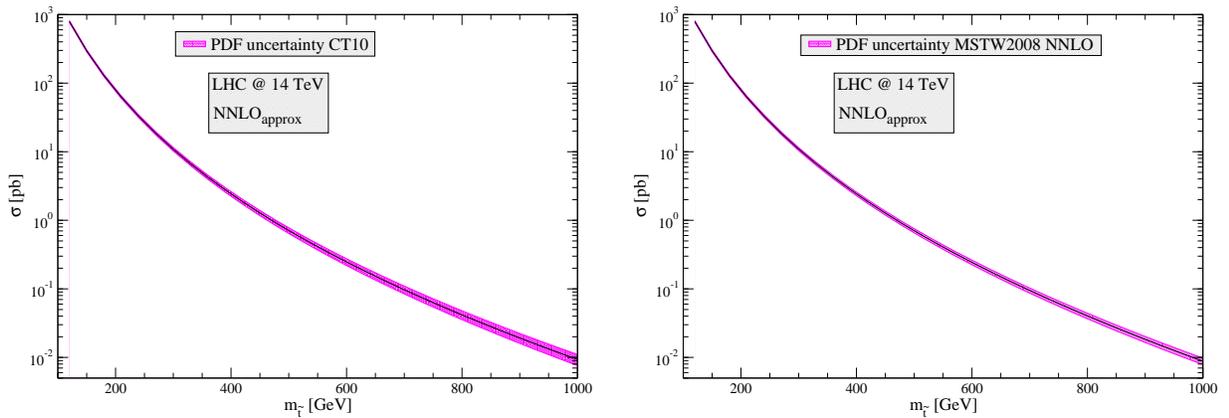

  \centering
	\vspace*{10mm}
  \scalebox{0.32}{\includegraphics{./pics/lhc14pdfuncertaintyct10.eps}}
  \hspace*{3mm}
  \scalebox{0.32}{\includegraphics{./pics/lhc14pdfuncertaintymstw.eps}}
  \caption{\small{PDF uncertainty of the total NNLO cross section for the two PDF sets
	CT10~\cite{Lai:2010vv} (left figure) 
	and MSTW2008 NNLO~\cite{Martin:2009iq} (right figure)
	at the LHC (14\TeV).}}
  \label{fig:tothadpdfunc}
\end{figure}
Combining theoretical uncertainty and PDF error one obtains
\begin{eqnarray}
\sigma_{\text{NNLO}} &=& 10.90\pb\enspace {}^{+0.01}_{-0.18}\pb\enspace
    (\text{scale})\enspace{}^{+0.55}_{-0.55}\pb\enspace(\text{MSTW2008 NNLO})\\[1mm]
\sigma_{\text{NNLO}} &=& 10.86\pb\enspace {}^{+0.01}_{-0.19}\pb\enspace
    (\text{scale})\enspace{}^{+0.65}_{-0.64}\pb\enspace(\text{CT10}).
\end{eqnarray}
 
\subsection{Mass exclusion limits}
The approximated NNLO contributions enlarge the stop-antistop production
cross section by $\approx10 - 20\%$, depending on the hadron collider,
its center of mass energy, and the stop mass.
This could be converted into larger exclusion limits for the mass
of the stop squark.
A stop with a mass $\mstop[1]=120\GeV$ has a NLO production 
cross section of $5.05\pb$,
the same cross section corresponds at NNLO to a stop with a mass 
of $123.5\GeV$. At higher stop masses, the exclusion limit is even further enhanced:
$0.164\pb$ corresponds at NLO to a stop with a mass of $210\GeV$, 
but at NNLO to a mass of $215\GeV$.
At the LHC ($14\TeV$) one would have a similar situation.
An NLO cross section of $750\pb$ corresponds to $\mstop[1] = 120\GeV$,
but to $\mstop[1] = 121.5\GeV$ at NNLO.
And if the stops are heavier,  a cross section of $10\pb$ is related to 
$\mstop[1] = 300\GeV$ at NLO, but to $\mstop[1] = 305.5\GeV$ at NNLO.   
     
\section{Conclusion and Summary}
In this paper, I computed the NNLO threshold contributions 
including Coulomb corrections for 
stop-antistop production at hadron colliders.

\begin{itemize}
\item I presented analytical formulas for the threshold expansion of
	 the NNLO scaling function using resummation techniques for the 
	 scale independent scaling function and RGE techniques for the 
	 scale-dependent scaling functions. 
\item After convolution with suitable PDF sets,
	 the NNLO corrections are found to be about $20\%$ for the Tevatron
	 and $10-20\%$ for the LHC compared to the hadronic NLO cross section.
	The PDF sets Cteq6.6~\cite{Nadolsky:2008zw},
	MSTW 2008 NNLO~\cite{Martin:2009iq},
     and CT10~\cite{Lai:2010vv} show only small differences
	in the total cross section, whereas the values obtained with 
	the PDF set ABKM09 NNLO (5 flavours)~\cite{Alekhin:2009ni}
	differ by $10-35\%$ due to differences in the gluon PDF.
\item I calculated the exact scale dependence and I found a remarkable 
	 stabilisation of the cross section under scale variation.
	 For my example point, the theoretical error is reduced from
	 $12\%$ at NLO to better than $2\%$ at NNLO.
\item I discussed three types of errors: systematic theoretical
	 errors, uncertainties due to scale variation and PDF errors.
	 The systematic error was estimated to be about $3-6\%$,
	 the scale uncertainty to be about $2\%$ or better,
	 and the PDF error $3-18\%$ depending on the stop mass
	 and the PDF set used.
\item Finally, I demonstrated how NNLO cross sections could enlarge
	 exclusion limits. The improvement of the lower exclusion limit 
	 were about a few $\GeV$.
\end{itemize}

\section*{Acknowledgments}
I would like to thank P. Uwer and A. Kulesza, W. Porod, and S. Uccirati
for useful discussions and 
S. Moch and M. Kr\"amer for reading the manuscript and giving helpful
comments.
This work is supported in part by the Helmholtz Alliance 
{\it ``Physics at the Terascale''} (HA-101)
and by the research training group GRK 1147 of the Deutsche 
Forschungs\-gemein\-schaft. 

\appendix
\section{Useful Formulas}
\subsection{Mellin Transformations}
\label{subsec:mellininversion}
\begin{eqnarray}
\label{eq:inveins}
\int_0^1 \dif\rho \rho^{N-1} \beta^3\lnbeta &=&
 -\frac{\sqrt{\pi}}{8} \Big[-8 + 6\*\lnzwei + 3\*\lnNt \Big]\frac{1}{N^{5/2}}
   \big(1 + \mathcal{O}\left(1/N\right)\big)\\
\label{eq:invzwei}
\int_0^1 \dif\rho \rho^{N-1} \beta^3\lnbeta[2] &=&
\frac{\sqrt{\pi}}{32} \Big[
3\*\pi^2 +16-64\*\lnzwei + 24\*\lnzwei[2]
+8\*\big(-4+3\*\lnzwei\big)\lnNt+6\*\lnNt[2]
\Big]\notag\\[-3mm]
&&\hspace*{10mm}
\times\frac{1}{N^{5/2}}
\big(1 + \mathcal{O}\left(1/N\right)\big)\\
\label{eq:invdrei}
\int_0^1 \dif\rho \rho^{N-1} \beta^3\lnbeta[3] &=&
-\frac{3\sqrt{\pi}}{64} \Big[
32\lnzwei-64\lnzwei[2]+16\lnzwei[3]-8\pi^2+6\*\pi^2\*\lnzwei + 28\*\zeta_3 \notag\\[-3mm]
&&\hspace*{15mm}
+\big(16 + 3\*\pi^2-64\*\lnzwei+24\*\lnzwei[2]\big)\*\lnNt \notag\\[1mm]
&&\hspace*{15mm}
+2\*\big(-8+6\*\lnzwei\big)\*\lnNt[2]
+2\*\lnNt[3]
\Big]
\frac{1}{N^{5/2}}
\big(1 + \mathcal{O}\left(1/N\right)\big)\\
\label{eq:invvier}
\int_0^1 \dif\rho \rho^{N-1} \beta^3\lnbeta[4] &=&
\frac{\sqrt{\pi}}{256} \Big[
21\pi^4+12\*\pi^2\*\big(8-32\*\lnzwei+12\*\lnzwei[2]\big)\notag\\[-3mm]
&&\hspace*{10mm}
+4\*\big(192\*\lnzwei[2]-256\*\lnzwei[3]+48\*\lnzwei[4]-448\*\zeta_3
+336\*\zeta_3\lnzwei\big)\notag\\[2mm]
&&\hspace*{10mm}
+24\*\big(-8\*\pi^2+6\*\pi^2\*\lnzwei+32\*\lnzwei-64\lnzwei[2]+16\*\lnzwei[3]
+28\*\zeta_3\big)\lnNt\notag\\[2mm]
&&\hspace*{10mm}
+12\*\big(16+3\*\pi^2-64\*\lnzwei+24\*\lnzwei[2]\big)\lnNt[2]\notag\\[1mm]
&&\hspace*{10mm}
+16\*\big(-8+6\*\lnzwei\big)\lnNt[3] + 12\*\lnNt[4]
\Big]
\frac{1}{N^{5/2}}
\big(1 + \mathcal{O}\left(1/N\right)\big)\\
\label{eq:invceins}
\int_0^1 \dif\rho \rho^{N-1} \beta^2\lnbeta &=&
-\frac{1}{2}\lnNt
\frac{1}{N^{2}}
\big(1 + \mathcal{O}\left(1/N\right)\big)\\
\label{eq:invczwei}
\int_0^1 \dif\rho \rho^{N-1} \beta^2\lnbeta[2] &=&
\frac{1}{24}\Big[\pi^2-12\*\lnNt+6\*\lnNt[2]\Big]
\frac{1}{N^{2}}
\big(1 + \mathcal{O}\left(1/N\right)\big)
\end{eqnarray}
with $\lnNt = \ln N + \gamma_e$.
A detailed overview of Mellin transformations is given in Ref.~\cite{Blumlein:2009ta}.
\newpage
\subsection{Resummation}
\label{subsec:resum}
For the sake of clarity, I specify the function $g^0_{q\bar{q},8}$
I have used in my resummation.
The function $g^0_{q\bar{q},8}$ differs a little bit from the function
given in Ref.~\cite{Moch:2008qy} due to Mellin transformation of functions
of the type $\beta^3\log^k(\beta)$ instead of $\beta\log^k(\beta)$, 
$0\leq k \leq 4$, as it is in the case of $t\bar{t}$-production.
\begin{eqnarray}
\label{eq:g0qq}
g^0_{q\bar{q},8} &=&
1+{\as}\* \Big[ {  \Cf}\* \big( -64+8\*{\gamma_e}^{2}+4\*{\pi }^{2
}-32\*\lnzwei +8\* \lnzwei[2] \big) 
+{  \Ca}\* \left( -8+4\*\gamma_e-4\*\ln 2  \right) 
+4\*{  C^{(1)}_{q\bar{q}}}\notag\\[1mm]
&&\hspace*{10mm}
+2\*  \Cf\* \left( 16-8\*\gamma_e \right) \ln 2  \Big] \notag\\[1mm]
 &&\hspace*{3mm}
+ \as^{2} \biggl[ 
 \Cf\*  C^{(1)}_{q\bar{q}}\* \biggl( -{\frac {3328}{9}}+32\*{\gamma_e}^{2}
+16\*{\pi }^{2}+512\*\ln 2  -64\*\gamma_e\*\ln 2  
-256 \*\lnzwei[{2}] \biggr) \notag\\[1mm]
&&\hspace*{10mm}
+\Cf\*\nf\* \biggl( {\frac {30592}{81}}-{\frac {448}{9}}\*{  \zeta_3}-{\frac {
224}{27}}\*\gamma_e-{\frac {80}{9}}\*\gamma_e^{2}-{\frac {32}{9}}\*
\gamma_e^{3}-{\frac {56}{3}}\*{\pi }^{2}-{\frac {14272}{27}}\*\ln 2\notag\\[1mm]
&&\hspace*{20mm}
+{\frac {160}{9}}\*\gamma_e\*\ln 2  +{
\frac {32}{3}}\*\gamma_e^{2}\ln 2  +16\*{\pi }^{2}\ln 2  
+{\frac {2944}{9}}\* \lnzwei[2] 
-{\frac {32}{3}}\*\gamma_e\* \lnzwei[2] 
-{\frac {832}{9}}\* \lnzwei[3]
\biggr) \notag\\[1mm]
&&\hspace*{10mm}
+\Cf^{2} \biggl( -{\frac {335872}{27}}
+{\frac {14336}{3}}\*{  \zeta_3}-512\*\gamma_e^{2}+32\*\gamma_e^{4}
+{\frac{4352}{9}}\*{\pi }^{2}+32\*{\pi }^{2}\gamma_e^{2}+24\*{\pi }^{4}
\notag\\[1mm]
&&\hspace*{20mm}
+{\frac {57344}{3}}\*\lnzwei
 -5376\*{  \zeta_3}\*\lnzwei +1024\*\gamma_e\*\lnzwei
-128\*\gamma_e^{3}\lnzwei -1024\*{\pi }^{2}\lnzwei 
\notag\\[1mm]
&&\hspace*{20mm}
-64\*{\pi }^{2}\gamma_e\*\lnzwei
-15872\* \lnzwei[2] +192\*\gamma_e^{2} \lnzwei[2]
+608\*{\pi }^{2} \lnzwei[2] +9216\* \lnzwei[3]
\notag\\[1mm]
&&\hspace*{20mm}
-128\*\gamma_e\* \lnzwei[3]-2560\* \lnzwei[4]
\biggr) \notag\\[1mm]
&&\hspace*{10mm}
+ \Ca\*  C^{(1)}_{q\bar{q}}\* \left( -{\frac {128}{3}}
+16\*\gamma_e+32\*\lnzwei 
\right) \notag\\[1mm]
&&\hspace*{10mm}
+\Ca\*\nf\*
 \left( \frac {512}{9}-{\frac {88}{9}}\*\gamma_e-\frac{8}{3}\*\gamma_e^{2}
-\frac{4}{3}\*{\pi }^{2}-\frac {464}{9}\*\lnzwei
+\frac{16}{3}\*\gamma_e\*\lnzwei +\frac {64}{3}\* \lnzwei[2]
\right) \notag\\[1mm]
&&\hspace*{10mm}
+\Ca\*\Cf\* \biggl( -\frac {250432}{81}
+{\frac {7840}{9}}\*{  \zeta_3}-{\frac {5296}{27}}\*\gamma_e-56\*\gamma_e\*
{  \zeta_3}-{\frac {40}{9}}\*\gamma_e^{2}+{\frac {464}{9}}\*\gamma_e^{3}+
{\frac {5188}{27}}\*{\pi }^{2}
\notag\\[1mm]
&&\hspace*{20mm}
+16\*{\pi }^{2}\gamma_e
-\frac{8}{3}\*{\pi }^{2}\gamma_e^{2}-\frac{4}{3}{\pi }^{4}
+\frac {132256}{27}\*\lnzwei-112\*\zeta_3\*\lnzwei
+\frac {80}{9}\*\gamma_e\*\lnzwei
\notag\\[1mm]
&&\hspace*{20mm}
-\frac {464}{3}\*\gamma_e^{2}\lnzwei 
-{\frac {728}{3}}\*{\pi }^{2}\lnzwei +\frac{16}{3}\*{\pi }^{2}\gamma_e\*\lnzwei 
-{\frac {33088}{9}}\* \lnzwei[2] +{\frac {464}{3}}\*\gamma_e\* \lnzwei[2]
\notag\\[1mm]
&&\hspace*{20mm}
+{\frac {64}{3}}\*{\pi }^{2} \lnzwei[2]
+{\frac {12064}{9}}\*\lnzwei[3] 
\biggr)  \notag\\[1mm]
&&\hspace*{10mm}
+\Ca^{2}
 \biggl( -{\frac {2816}{9}}+{\frac {172}{9}}\*\gamma_e
+{\frac {68}{3}}\*\gamma_e^{2}+{\frac {134}{9}}\*{\pi }^{2}
-\frac{4}{3}\*{\pi }^{2}\gamma_e+{\frac {3608}{9}}\*\lnzwei
-{\frac {136}{3}}\*\gamma_e\*\lnzwei 
\notag\\[1mm]
&&\hspace*{20mm}
-\frac{8}{3}\*{\pi }^{2}\lnzwei -{\frac {544}{3}}\* \lnzwei[2]
-\frac{64}{3}\* \zeta_3 + 16\*\zeta_3\lnzwei+8\*\gamma_e\*\zeta_3
\biggr) \quad 
\notag\\[1mm]
&&\hspace*{10mm}
+  C^{(2)}_{q\bar{q}}
\biggr]\quad .
\end{eqnarray}
$\gamma_e$ denotes the Euler constant and $\zeta_3$ is the Riemannian
Zeta function $\zeta(x)$ evaluated at $x=3$. 
\section{Numerical Results}
\label{subsec:tables}
\begin{table}[tbh]
\centering
 \begin{tabular}{c|rrr|rrr|rrr}
\toprule
$\mstop$ &\multicolumn{3}{c|}{$\sigma(\text{LO}) [\pb]$}&
       \multicolumn{3}{c|}{$\sigma(\text{NLO}) [\pb]$}&
       \multicolumn{3}{c}{$\sigma(\text{NNLO}) [\pb]$}\\
$[\GeV]$& $x = \half$& $x = 1$& $x = 2$& $x = \half$& $x = 1$& $x = 2$& 
          $x = \half$& $x = 1$& $x = 2$\\
\hline
\multicolumn{10}{c}{Cteq6.6}\\
\hline
 120 &  579.745 &  461.516 &  372.340 &  842.538 &  741.438 &  652.012 &  762.236 &  786.927 &  785.150 \\ 
 180 &   97.170 &   75.718 &   60.001 &  133.421 &  118.340 &  103.980 &  124.929 &  127.242 &  126.175 \\ 
 240 &   25.299 &   19.463 &   15.254 &   33.578 &   29.929 &   26.279 &   32.176 &   32.490 &   32.086 \\ 
 300 &    8.466 &    6.455 &    5.020 &   10.976 &    9.817 &    8.615 &   10.702 &   10.739 &   10.573 \\ 
 360 &    3.335 &    2.526 &    1.953 &    4.250 &    3.810 &    3.342 &    4.201 &    4.196 &    4.121 \\ 
 480 &    0.710 &    0.532 &    0.408 &    0.883 &    0.794 &    0.696 &    0.892 &    0.884 &    0.865 \\ 
 600 &    0.198 &    0.147 &    0.112 &    0.242 &    0.218 &    0.191 &    0.249 &    0.245 &    0.239 \\ 

\hline
\multicolumn{10}{c}{MSTW 2008 NNLO}\\
\hline
 120 &  587.154 &  468.209 &  378.090 &  849.643 &  749.028 &  659.621 &  769.079 &  793.744 &  791.998 \\ 
 180 &   99.033 &   77.177 &   61.136 &  135.263 &  120.146 &  105.649 &  126.755 &  129.046 &  127.921 \\ 
 240 &   25.825 &   19.848 &   15.538 &   34.083 &   30.413 &   26.712 &   32.698 &   32.996 &   32.566 \\ 
 300 &    8.632 &    6.571 &    5.101 &   11.127 &    9.960 &    8.740 &   10.867 &   10.896 &   10.718 \\ 
 360 &    3.390 &    2.562 &    1.977 &    4.295 &    3.854 &    3.379 &    4.255 &    4.245 &    4.165 \\ 
 480 &    0.714 &    0.534 &    0.408 &    0.883 &    0.795 &    0.696 &    0.895 &    0.886 &    0.866 \\ 
 600 &    0.197 &    0.146 &    0.111 &    0.239 &    0.216 &    0.189 &    0.247 &    0.243 &    0.236 \\ 

\hline
\multicolumn{10}{c}{CT10}\\
\hline
 120 &  584.754 &  465.703 &  375.794 &  849.755 &  747.907 &  657.825 &  768.385 &  793.474 &  791.811 \\ 
 180 &   98.265 &   76.560 &   60.641 &  134.841 &  119.609 &  105.071 &  126.243 &  128.588 &  127.483 \\ 
 240 &   25.605 &   19.688 &   15.415 &   33.955 &   30.263 &   26.555 &   32.541 &   32.854 &   32.425 \\ 
 300 &    8.568 &    6.528 &    5.070 &   11.098 &    9.923 &    8.699 &   10.824 &   10.858 &   10.679 \\ 
 360 &    3.374 &    2.553 &    1.970 &    4.294 &    3.848 &    3.371 &    4.248 &    4.239 &    4.158 \\ 
 480 &    0.717 &    0.537 &    0.410 &    0.890 &    0.800 &    0.700 &    0.900 &    0.891 &    0.870 \\ 
 600 &    0.199 &    0.148 &    0.112 &    0.244 &    0.219 &    0.191 &    0.250 &    0.246 &    0.240 \\ 

\hline
\multicolumn{10}{c}{ABKM 09 NNLO (5 flv)}\\
\hline
 120 &  539.054 &  428.594 &  345.556 &  764.660 &  677.764 &  598.338 &  700.513 &  720.046 &  718.084 \\ 
 180 &   87.613 &   68.261 &   54.112 &  117.982 &  105.305 &   92.854 &  111.788 &  113.439 &  112.487 \\ 
 240 &   22.132 &   17.046 &   13.380 &   28.906 &   25.912 &   22.833 &   28.027 &   28.207 &   27.868 \\ 
 300 &    7.190 &    5.498 &    4.288 &    9.196 &    8.270 &    7.284 &    9.075 &    9.079 &    8.947 \\ 
 360 &    2.752 &    2.093 &    1.626 &    3.464 &    3.123 &    2.751 &    3.468 &    3.454 &    3.397 \\ 
 480 &    0.554 &    0.419 &    0.323 &    0.682 &    0.617 &    0.544 &    0.698 &    0.690 &    0.677 \\ 
 600 &    0.146 &    0.110 &    0.085 &    0.177 &    0.161 &    0.142 &    0.185 &    0.182 &    0.178 \\ 

\bottomrule
 \end{tabular}
    \caption{\small
      \label{tab:xsecvalues}
      Numerical values for the stop pair-production cross section
      at LHC with $\sqrt{s}=14\TeV$ and the PDF sets 
     Cteq6.6~\cite{Nadolsky:2008zw},
	MSTW 2008 NNLO~\cite{Martin:2009iq},
	ABKM09 NNLO (5 flavours)~\cite{Alekhin:2009ni},
	and CT10~\cite{Lai:2010vv}.
      The QCD predictions are given at LO, NLO, and NNLO accuracy and
      for different stop masses and scales $x = \mu/\mstop$.}
\end{table}

\begin{table}
\centering
 \begin{tabular}{c|rrr|rrr|rrr}
\toprule
$\mstop$ &\multicolumn{3}{c|}{$\sigma(\text{LO}) [\pb]$}&
       \multicolumn{3}{c|}{$\sigma(\text{NLO}) [\pb]$}&
       \multicolumn{3}{c}{$\sigma(\text{NNLO}) [\pb]$}\\
$[\GeV]$& $x = \half$& $x = 1$& $x = 2$& $x = \half$& $x = 1$& $x = 2$& 
          $x = \half$& $x = 1$& $x = 2$\\
\hline
\multicolumn{10}{c}{Cteq6.6}\\
\hline
 120 &  134.391 &  101.197 &   77.851 &  184.136 &  160.840 &  138.709 &  173.624 &  176.078 &  173.026 \\ 
 180 &   18.038 &   13.341 &   10.106 &   23.542 &   20.725 &   17.855 &   23.084 &   23.068 &   22.511 \\ 
 240 &    3.881 &    2.839 &    2.130 &    4.925 &    4.354 &    3.748 &    4.957 &    4.907 &    4.765 \\ 
 300 &    1.091 &    0.792 &    0.590 &    1.359 &    1.205 &    1.036 &    1.395 &    1.372 &    1.327 \\ 
 360 &    0.364 &    0.263 &    0.195 &    0.448 &    0.398 &    0.342 &    0.468 &    0.457 &    0.441 \\ 
 420 &    0.137 &    0.098 &    0.072 &    0.168 &    0.149 &    0.128 &    0.177 &    0.172 &    0.165 \\ 

\hline
\multicolumn{10}{c}{MSTW 2008 NNLO}\\
\hline
 120 &  137.257 &  103.302 &   79.392 &  186.837 &  163.495 &  141.098 &  176.403 &  178.825 &  175.655 \\ 
 180 &   18.377 &   13.561 &   10.249 &   23.825 &   20.998 &   18.086 &   23.416 &   23.378 &   22.789 \\ 
 240 &    3.918 &    2.857 &    2.137 &    4.941 &    4.371 &    3.759 &    4.991 &    4.933 &    4.782 \\ 
 300 &    1.087 &    0.786 &    0.584 &    1.346 &    1.194 &    1.025 &    1.388 &    1.362 &    1.315 \\ 
 360 &    0.357 &    0.257 &    0.190 &    0.437 &    0.388 &    0.333 &    0.458 &    0.447 &    0.430 \\ 
 420 &    0.132 &    0.094 &    0.069 &    0.160 &    0.142 &    0.122 &    0.170 &    0.165 &    0.159 \\ 

\hline
\multicolumn{10}{c}{CT10}\\
\hline
 120 &  136.074 &  102.422 &   78.752 &  186.261 &  162.710 &  140.298 &  175.639 &  178.121 &  175.005 \\ 
 180 &   18.262 &   13.496 &   10.211 &   23.805 &   20.956 &   18.042 &   23.354 &   23.332 &   22.753 \\ 
 240 &    3.922 &    2.866 &    2.146 &    4.971 &    4.394 &    3.777 &    5.008 &    4.955 &    4.805 \\ 
 300 &    1.100 &    0.797 &    0.592 &    1.368 &    1.212 &    1.040 &    1.406 &    1.381 &    1.333 \\ 
 360 &    0.366 &    0.263 &    0.195 &    0.450 &    0.399 &    0.342 &    0.470 &    0.459 &    0.441 \\ 
 420 &    0.137 &    0.098 &    0.072 &    0.167 &    0.148 &    0.127 &    0.177 &    0.172 &    0.165 \\ 

\hline
\multicolumn{10}{c}{ABKM 09 NNLO (5 flv)}\\
\hline
 120 &  117.501 &   88.531 &   68.185 &  157.745 &  138.823 &  120.224 &  151.054 &  152.485 &  149.896 \\ 
 180 &   14.825 &   11.007 &    8.373 &   19.061 &   16.894 &   14.626 &   18.991 &   18.902 &   18.474 \\ 
 240 &    3.004 &    2.215 &    1.674 &    3.765 &    3.353 &    2.904 &    3.856 &    3.804 &    3.704 \\ 
 300 &    0.797 &    0.586 &    0.441 &    0.981 &    0.877 &    0.761 &    1.027 &    1.006 &    0.978 \\ 
 360 &    0.252 &    0.185 &    0.139 &    0.306 &    0.274 &    0.238 &    0.326 &    0.318 &    0.308 \\ 
 420 &    0.090 &    0.066 &    0.049 &    0.108 &    0.097 &    0.084 &    0.117 &    0.113 &    0.110 \\ 

\bottomrule
 \end{tabular}
    \caption{\small
      \label{tab:xsecvalueslhc07}
      Numerical values for the stop pair-production cross section
      at LHC with $\sqrt{s}=7\TeV$ and the same PDF sets as in 
	Tab.~\ref{tab:xsecvalues}.
      The QCD predictions are given at LO, NLO, and NNLO accuracy and
      for different stop masses and scales $x = \mu/\mstop$.}
\end{table}

\begin{table}
\centering
 \begin{tabular}{c|rrr|rrr|rrr}
\toprule
$\mstop$ &\multicolumn{3}{c|}{$\sigma(\text{LO}) [\pb]$}&
       \multicolumn{3}{c|}{$\sigma(\text{NLO}) [\pb]$}&
       \multicolumn{3}{c}{$\sigma(\text{NNLO}) [\pb]$}\\
$[\GeV]$& $x = \half$& $x = 1$& $x = 2$& $x = \half$& $x = 1$& $x = 2$& 
          $x = \half$& $x = 1$& $x = 2$\\
\hline
\multicolumn{10}{c}{Cteq6.6}\\
\hline
 120 &    5.525 &    3.850 &    2.779 &    5.782 &    5.246 &    4.509 &    6.453 &    6.144 &    5.780 \\ 
 150 &    1.570 &    1.095 &    0.790 &    1.572 &    1.448 &    1.254 &    1.793 &    1.695 &    1.589 \\ 
 180 &    0.525 &    0.366 &    0.264 &    0.519 &    0.480 &    0.417 &    0.597 &    0.562 &    0.524 \\ 
 210 &    0.194 &    0.135 &    0.097 &    0.192 &    0.178 &    0.154 &    0.222 &    0.208 &    0.193 \\ 
 240 &    0.076 &    0.053 &    0.038 &    0.076 &    0.070 &    0.061 &    0.088 &    0.082 &    0.076 \\ 
 270 &    0.031 &    0.021 &    0.015 &    0.032 &    0.029 &    0.025 &    0.036 &    0.034 &    0.031 \\ 
 300 &    0.013 &    0.009 &    0.006 &    0.013 &    0.012 &    0.010 &    0.015 &    0.014 &    0.013 \\ 

\hline
\multicolumn{10}{c}{MSTW 2008 NNLO}\\
\hline
 120 &    5.336 &    3.712 &    2.677 &    5.562 &    5.047 &    4.336 &    6.242 &    5.930 &    5.568 \\ 
 150 &    1.492 &    1.040 &    0.751 &    1.482 &    1.368 &    1.186 &    1.701 &    1.605 &    1.503 \\ 
 180 &    0.493 &    0.344 &    0.248 &    0.482 &    0.448 &    0.390 &    0.558 &    0.524 &    0.489 \\ 
 210 &    0.180 &    0.126 &    0.090 &    0.177 &    0.164 &    0.143 &    0.205 &    0.191 &    0.178 \\ 
 240 &    0.070 &    0.049 &    0.035 &    0.069 &    0.064 &    0.056 &    0.080 &    0.075 &    0.069 \\ 
 270 &    0.028 &    0.020 &    0.014 &    0.029 &    0.026 &    0.023 &    0.033 &    0.031 &    0.028 \\ 
 300 &    0.012 &    0.008 &    0.006 &    0.012 &    0.011 &    0.009 &    0.014 &    0.013 &    0.012 \\ 

\hline
\multicolumn{10}{c}{CT10}\\
\hline
 120 &    5.546 &    3.861 &    2.784 &    5.793 &    5.256 &    4.516 &    6.479 &    6.164 &    5.795 \\ 
 150 &    1.570 &    1.094 &    0.789 &    1.567 &    1.444 &    1.251 &    1.792 &    1.693 &    1.586 \\ 
 180 &    0.524 &    0.365 &    0.263 &    0.515 &    0.478 &    0.415 &    0.595 &    0.559 &    0.521 \\ 
 210 &    0.193 &    0.134 &    0.096 &    0.190 &    0.176 &    0.153 &    0.220 &    0.206 &    0.191 \\ 
 240 &    0.075 &    0.052 &    0.037 &    0.075 &    0.070 &    0.060 &    0.087 &    0.081 &    0.075 \\ 
 270 &    0.031 &    0.021 &    0.015 &    0.031 &    0.029 &    0.025 &    0.036 &    0.033 &    0.031 \\ 
 300 &    0.013 &    0.009 &    0.006 &    0.013 &    0.012 &    0.010 &    0.015 &    0.014 &    0.013 \\ 

\hline
\multicolumn{10}{c}{ABKM 09 NNLO (5 flv)}\\
\hline
 120 &    4.478 &    3.207 &    2.371 &    4.374 &    4.095 &    3.617 &    5.023 &    4.773 &    4.532 \\ 
 150 &    1.312 &    0.942 &    0.698 &    1.222 &    1.165 &    1.040 &    1.423 &    1.348 &    1.278 \\ 
 180 &    0.453 &    0.325 &    0.240 &    0.419 &    0.401 &    0.359 &    0.488 &    0.461 &    0.435 \\ 
 210 &    0.172 &    0.123 &    0.090 &    0.161 &    0.153 &    0.137 &    0.186 &    0.175 &    0.165 \\ 
 240 &    0.069 &    0.049 &    0.036 &    0.066 &    0.062 &    0.055 &    0.076 &    0.071 &    0.067 \\ 
 270 &    0.028 &    0.020 &    0.015 &    0.028 &    0.026 &    0.023 &    0.032 &    0.030 &    0.028 \\ 
 300 &    0.012 &    0.008 &    0.006 &    0.012 &    0.011 &    0.010 &    0.014 &    0.013 &    0.012 \\ 

\bottomrule
 \end{tabular}
    \caption{\small
      \label{tab:xsecvaluesteva}
      Numerical values for the stop pair-production cross section
      at the Tevatron with $\sqrt{s}=1.96\TeV$ and the same PDF sets as in 
	Tab.~\ref{tab:xsecvalues}.
      The QCD predictions are given at LO, NLO, and NNLO accuracy and
      for different stop masses and scales $x = \mu/\mstop$.}
\end{table}


\end{document}